\documentclass[11pt,a4paper]{article}
\pdfoutput=1
\usepackage{amsmath}                                       
\usepackage{amsthm}                                        
\usepackage{amssymb}                                       
\usepackage[round]{natbib}                  			
\setcitestyle{aysep={}}
\usepackage[center]{caption}
\usepackage[spanish,es-nodecimaldot,english]{babel}
\usepackage[latin1]{inputenc}
\usepackage[top=1.5in,bottom=1.5in,left=1in,right=1in,footskip=0.25in]{geometry}
\usepackage{graphicx}                                      
\usepackage{epstopdf}
\usepackage{colortbl}                                      
\usepackage{setspace}                                      
\usepackage{rotating}                                      
\usepackage{enumitem}                                      
\usepackage{booktabs}                                      
\usepackage[flushleft]{threeparttable}
\usepackage{multirow}                                      
\usepackage{tabularx}
\usepackage{dcolumn}
\usepackage{color}
\usepackage{eurosym}
\usepackage{url}
\definecolor{darkred}{rgb}{0.5,0,0}
\definecolor{darkblue}{rgb}{0,0,0.5}
\usepackage[colorlinks,breaklinks,bookmarksnumbered,bookmarksopenlevel=2,unicode]{hyperref}               
\hypersetup{
  colorlinks,
  citecolor=darkred,
  linkcolor=darkred,
  urlcolor=darkblue,
	bookmarksopen=true, bookmarksopenlevel=3
	}

\usepackage{float}
\usepackage{lscape}
\usepackage[centerlast]{subfigure}
\usepackage{setspace}
\usepackage{caption}
\usepackage{siunitx}

\begin{document}

\title{An internal fraud model for operational losses in retail banking}
\author{Roc\'io Paredes\thanks{Centrum Cat\'olica Graduate Business School (\href{mailto:rmparedes@pucp.edu.pe}{rocio.m.paredes@pucp.edu.pe}) and Pontificia Universidad Cat\'olica del Per\'u.} \and Marco Vega\thanks{Banco Central de Reserva del Per\'u (\href{mailto:marco.vega@bcrp.gob.pe}{marco.vega@bcrp.gob.pe}) and Pontificia Universidad Cat\'olica del Per\'u.}}
\date{This version: June 13, 2019}
\maketitle

\begin{abstract}
\normalsize{This paper develops a dynamic internal fraud model for operational losses in retail banking. It considers public operational losses arising from internal fraud in retail banking within a group of international banks. Additionally, the model takes into account internal factors such as the ethical quality of workers and the risk controls set by bank managers. The model is validated by measuring the impact of macroeconomic indicators such as GDP growth and the corruption perception upon the severity and frequency of losses implied by the model. In general,results show that internal fraud losses are pro-cyclical, and that country specific corruption perceptions positively affects internal fraud losses. Namely, when a country is perceived to be more corrupt, retail banking in that country will feature more severe internal fraud losses.}
\end{abstract}

\selectlanguage{english}


\newpage
\doublespacing

\section{Introduction}

This paper studies operational losses due to internal fraud in the retail segment of banks. Internal fraud is defined as operational losses due to acts that involve at least one internal party aimed at defrauding, misappropriating property, or circumventing regulations, the law, or company policies \citep{comitato2006international}.  As \citet{kochan2013} suggests, internal fraud is one of the fastest-growing and most complex criminal threats to financial organizations.  This type of threat from insiders takes various forms because fraud can occur at any level of the administrative ladder, from junior employees up to chief executives. Internal fraud events are due to factors such as worker compensation, culture, or macroeconomic conditions \citep{jarrow2008}

Retail banking is a traditional, universal type of banking involving payment services (debit cards), short-term unsecured loans (credit cards), money management facilities (current accounts), savings, loans, and mortgages.  According to \citet{orx2012}, retail banking experiences the larger number of operational loss events, 59\% for the period 2006-2010, and increasing to 65\% in 2011.  In addition, the gross losses in retail banking are the most severe of losses across business lines, representing up to 37\% of total losses by business line.

The nexus between internal fraud and retail banking has not been explored much in the literature. Given the importance of the retail segment in the banking business, it is important to shed some light in the topic by laying out a dynamic model of loss generation in the internal fraud - retail banking cell. The model is set to mimic the aggregate behaviour of these losses in the ORX database.\footnote{The Operational Riskdata eXchange Association (ORX) is a global data sharing association whose members comprise the biggest banks in the world.} In addition, other specific information about banks that participate in the ORX consortium is obtained from different outlets such as financial service authorities and their corporate Web pages. The model is validated by measuring how observed macroeconomic variables outside the model affect the loss severities and frequencies. The paper finds that there is a strong association between macro variables and fraud losses in retail banking.

The model can be used as an operational risk management tool. Scenario analysis and operational risk capital simulations are straightforward to perform. For example, the model can be used to evaluate different data aggregation techniques in dealing with consortium data.

To the best of our knowledge, the operational risk management literature has not provided a quantitative model for describing internal fraud in the financial sector; this study is the first to contribute in this direction.  The internal fraud model borrows insights from a number of disciplines such as corporate governance, behavioral economics, human resources, and operational risk.

\noindent \textbf{Related literature}

In the operational risk literature, there is a group of papers that build on quantitative models that explain the  outbreak of operational loss events. \citet{kuhn2003functional} studies models that generate operational losses in banks through network dynamics that lead to the occurrence of risk events in an environment where banks make efforts to mitigate operational losses.  \citet{leippold2005quantification} uses functional dependence modeling to extend the work of \citet{kuhn2003functional,kuhn2004adequate} by including fixed and stochastic costs that arise in case of operational risk events. Based on this strand of the literature, \citet{bardoscia2011dynamical} models the amount of operational losses recorded at a certain time in a certain process.

Our paper differs from the above studies in the scope of loss generation. Our model focuses on internal fraud in retail banking whereas the aforementioned studies are more general and thus take into account the network structure of operational losses.

Other papers like \citet{fragniere2010operations}, \citet{hatzakis2010operations} or \citet{weiss2011developing} introduce the quality and quantity of the workforce as a source of risk.  The importance of human capital in the operational loss process of financial institutions accords with the idea that the key process of a bank is the handling of information. Banking is known to be a knowledge-intensive business process \citep{weiss2011developing}.  This calls for a modeling approach that takes the quantity and quality of employees into account.  

Our paper shares the insights of the managerial approach in \citet{fragniere2010operations}; however, this latter study is focused on the planning of workforce capacity and do not touch on factors that generate operational losses due to internal fraud.

Our paper is also related to the literature that studies how macroeconomic and macro institutional factors affect operational losses such as \citet{allen2007cyclicality}, \citet{chernobai2011determinants}, \citet{moosa2011operational}, \citet{cope2012macroenvironmental}, \citet{li2015operational}, \citet{stewart2016bank}, \citet{abdymomunov2017us} and \citet{wagner2017operational}. Some of these papers stress the relationship between macroeconomic activity and operational losses. For instance, \citet{allen2007cyclicality} and \citet{wagner2017operational} provide evidence that loss severities are pro-cyclical. The evidence is not conclusive. \citet{abdymomunov2017us} finds evidence for greater aggregate losses in a downturn and \citet{moosa2011operational} finds that severity is positively associated with the unemployment rate. On the other hand, \citet{li2015operational} finds that loss severities are positively related to the size of the economy.

The above evidence takes operational losses in general. Closest to our paper is \citet{stewart2016bank} which provides evidence that bank fraud and economic activity are positively related. Also, \citet{cope2012macroenvironmental} finds that internal fraud losses are strongly and positively associated with the countries' legal frameworks that favor insider trading and are negatively associated with country-specific constraints on executive power in banks. Last, \citet{chernobai2011determinants} finds that the frequency of internal fraud losses in financial institutions depend positively on features such as market value, fast liability growth and financial distress while negatively related to default-risk and the age of firms.

Our paper also studies the relationship between macroeconomic variables and operational losses but different from the above papers, it does not use observed operational loss data but simulated internal fraud losses that mimic the ORX data. The association between those macro variables and operational losses serves to validate the model and provides first evidence pointing to a positive link between the corruption culture in a country and internal fraud in retail banking.

The rest of the paper is organized as follows, section \ref{sec:model} lays out the dynamic model for internal fraud losses, section \ref{sec:data} describes the publicly available data used in the paper, section \ref{sec:result} focuses on the calibration and regression results and section \ref{sec:conclu} concludes.

\section{A model for internal fraud events}\label{sec:model}
We setup a dynamic model for operational loss occurrence due to internal fraud in the retail segment of banks.  The model calibration aims to mimic the first moments of aggregate severity and frequency of operational losses drawn from ORX database reports. The model takes into account the specific factors that trigger internal fraud losses in each of the banks that took part of the ORX consortium during the period 2006-2010. 
\subsection{Setup}
The stochastic model aims to explain operational losses due to internal or insider fraud in the retail-banking context within each financial institution in terms of a set of conditioning factors.  The main equation in the model is given by
\begin{equation}
  l_{i,\tau}=\alpha_{i,0} \times ramp\left(\alpha_{i,1}+\alpha_{i,c}c_{i,\tau}+\alpha_{i,y} y_{i,\tau}  +\alpha_{i,q} q_{i,\tau}  +  \xi_{i,\tau}\right)  \label{eq:ramp}
\end{equation}

where $l_{i,\tau}$ stands for an internal fraud loss in retail banking at bank $i$ at moment $\tau$.  The Greek letter subscript $\tau$ denotes moments of time during a given year $t$.  In practice, it can represent days or hours within a year.  The variable $c_{i,\tau}$ is the investment or effort made by bank $i$ to avoid the operational loss, or it can measure the level of internal controls.  This variable can be measured as the share of monetary resources devoted to risk management and control and can be expressed as a percentage of operating costs.  Higher standards of internal risk controls ($c_{i,\tau}$ high) imply that the likelihood of operational loss events is reduced.  Internal fraud events are somewhat more controllable than losses originating from external sources \citep{chernobai2011determinants}.  This control aspect of operational losses is outlined, for example, in \citet{kochan2013}.

The amount $y_{i,\tau}$ represents the scale of production in the business line; for retail banking, it can represent the number of transactions with bank clients, or it can represent retail income.  A higher number of transactions imply that the likelihood of operational losses increases.  In the theory of people risk management, this scale level $y_{i,\tau}$ is a proxy for internal and external interactions, which give rise to operational losses due to fraud \citep{blacker2015people}: The bigger the scale of the business, the higher the number of interactions.  In an environment of increased employee interaction, fraud risks rise.

Variable $q_{i,\tau}$ measures the ethical quality of employees.  High internal ethical standards mean that losses due to internal fraud are less likely to occur.  The ethical quality of workers is different from the technical quality of workers, which is measured directly by worker productivity (e.g., gross income per worker).  Therefore, the quantity and quality of human capital proposed by \citet{fragniere2010operations} are key determinants of operational losses in retail banking.

From the variables explained so far, the volume of gross income $y_{i,\tau}$ is directly observable.  Information about these variables can be gathered from the annual reports of each of the banks in the ORX dataset.  On the other hand, the level of controls $c_{i,\tau}$ and the quality of employees $q_{i,\tau}$ are not directly observable.  Specific feedback equations are required to model both variables to elicit their unobserved values and see how they quantitatively affect the generation of losses through Eq. (\ref{eq:ramp}).

The last variable left to be explained in Eq. (\ref{eq:ramp}) is $\xi_{i,\tau}$.  This variable represents unknown factors or shocks that can potentially trigger losses.  This random variable is assumed to be autocorrelated and heteroskedastic.  The idea that loss shocks are autocorrelated and may exhibit volatility clustering is similar to what \citet{chernobai2008dynamics} and \citet{guegan2013using} suggested.  In particular
\begin{equation}
 \xi_{i,\tau}=\rho_i \xi_{i,\tau-1}+\sigma_{i,\tau} \mu_{i,\tau}, \hspace{1cm} \mu_{i,\tau} \sim N(0,1), \hspace{1cm} \sigma_{i,\tau}^{2}=  \beta_{0,i}+\beta_{1} \xi_{i,\tau}^{2} + \beta_2 \sigma_{i,\tau-1}^{2} \label{eq:shock}
\end{equation}

The coefficient $\rho_i \in [0,1]$ measures the level of autocorrelation or the persistent nature of shocks that may trigger losses.  The error term $\sigma_{i,\tau} \mu_{i,\tau}$ is heteroskedastic by virtue of the time-varying nature of the variance term.  The variance term $\sigma_{i,\tau}^{2}$, also known as conditional variance, depends on past shock realizations as well as past variance itself.\footnote{The way conditional variance behaves is called generalized autoregressive conditional heteroscedasticity (GARCH), as proposed in \citet{bollerslev1986generalized}.}

Eq. (\ref{eq:ramp}) also calls the function $ramp(.)$, which represents the mapping from operational loss factors to loss severities.  This function has the feature of generating zero losses most of the time and positive loss severities at other times.\footnote{\citet{kuhn2004adequate} and \citet{bardoscia2011dynamical} used the same type of function.}. Formally, the ramp function is defined by
\begin{equation*}
ramp(x) = \begin{cases}
x & \text{if $x \geq 0$}\\
0 & \text{of $x <  0$}
\end{cases}
\end{equation*}

To finish the description of Eq. (\ref{eq:ramp}), all the coefficients $(\alpha_{i,0},\alpha_{i,1},\alpha_{i,c},\alpha_{i,y},\alpha_{i,q})$ vary across banks, but they are constant through time.  This reflects the fact that operational loss occurrences are sensitive to each of the factors described, which are idiosyncratic for each bank. The levels of operational risk control ($c_{i,\tau}$), and the quality of the workforce ($q_{i,\tau}$) need to be modeled.


The level of operational risk controls ($c_{i,\tau}$) obey a feedback equation whereby the controls or efforts by risk managers to prevent or mitigate operational losses depend on the observable state of the system.  Control is a fundamental aspect of risk management, the actual ISO standard defines risk management as ``coordinated activities to direct and control an organization with regard to risk'' \citep{iso2009}.
%

In this study, the levels of controls are represented by a sufficient statistic denoted by $c_{i,\tau}$.  The learning or improved control process depends on the level of risk.  This idea is common in stochastic control environments and follows the example of \citet{cooke2005conceptual} who proposed a very general feedback model of operational risk.  Our paper incorporates this idea and is explicit about the level of risk that feeds back onto the control level.  The feedback control equation can be expressed as
\begin{equation}
  c_{i,\tau} = \frac{\rho_c (2c_i^{\ast})}{1+exp\left(\gamma_i(\frac{\widehat{L}_{i,\tau-1}}{\widehat{Y}_{i,\tau-1}}-\lambda_i)\right)} + (1-\rho_c)c_{i,\tau-1}   \label{eq:control}
\end{equation}
where $c_i^{\ast}$ stands for the optimal level of controls associated with a benchmark loss ratio $\lambda_i$  when the actual loss ratio is given by $\frac{\widehat{L}_{i,t}}{\widehat{Y}_{i,t}}$.  The control level at bank $i$ depends on the observed key risk performance given by the ratio of cumulative average observed losses over cumulative income.  If the observed loss ratio is beyond the desired level, with $\gamma_i<0$, control levels need to be adjusted upward.  The degree of the actual adjustment depends on the parameter $\rho_c \in [0,1]$.  The higher $\rho_c$, the quicker the control response is.  In the opposite case, when $\rho_c$ is small, the control level is mostly governed by its previous value.

In Eq. (\ref{eq:control}), $\widehat{L}_{i,\tau-1}$ stands for the cumulative average observed losses up to the previous time, while $\widehat{Y}_{i,\tau-1}$ corresponds to cumulative gross retail income obtained in the same period.  Capital letters stand for average quantities, while the circumflex denotes that the variable is the observable counterpart of an unobservable underlying variable.  In the case of operational losses, this distinction is important.  An observed loss amount in a bank $i$ at time $\tau$ is $\widehat{l}_{i,\tau}$ whereas the true loss is $l_{i,\tau}$. Eq. (\ref{eq:control}) means that banks, which experience a history of large losses relative to other banks, will learn from the incidents and therefore increase their controls to levels above average.  This idea is also suggested in \citet{lukic2013individual}.

On the other hand, the quality of the workforce ($q_{i,\tau}$), which refers to ethical traits that drive workers' behavior toward the bank is a measure of the propensity to commit fraud.  For example, an employee can be extremely knowledgeable of internal processes at the bank and so be highly productive, but good knowledge of internal processes may make it easy to commit fraud \citep{cummings2012insider}.
The equation that describes the ethical quality of workers is
\begin{equation}
  q_{i,\tau} = \frac{2\bar{Q}\rho_q}{1 + \exp\left(\delta_i(a_{i,\tau} - \bar{A}_\tau)(e_{i,\tau} - \bar{E}_\tau) \right)}+(1-\rho_q ) q_{i,\tau-1}   \label{eq:quality}
\end{equation}
where $a_{i,\tau}$ stands for measured technical quality (labor productivity), $\bar{A}_\tau$ is the cross-bank average labor productivity, $e_{i,\tau}$ is the number of employees per branch at bank $i$, and $\bar{E}_\tau$ is the cross-bank average of employees per branch. Given that $\delta_i>0$, the sign of the impact effect of an increase in technical quality is given by
\begin{equation*}
\frac{\partial q_{i,\tau}}{\partial a_{i,\tau}} = \begin{cases}
< 0  & \text{if $e_{i,\tau} > \bar{E}_\tau$}\\
\geq & \text{if $e_{i,\tau} \leq \bar{E}_\tau$}
\end{cases}
\end{equation*}

When there are few workers, increasing productivity is more likely associated with high ethical quality because it is easier for banks to screen workers before and after recruitment. When the number of workers is high, the workforce screening process is weaker. Due to the symmetry of Eq. (\ref{eq:quality}), it is also true that
\begin{equation*}
\frac{\partial q_{i,\tau}}{\partial e_{i,\tau}} = \begin{cases}
< 0  & \text{if $a_{i,\tau} > \bar{A}_\tau$}\\
\geq & \text{if $a_{i,\tau} \leq \bar{A}_\tau$}
\end{cases}
\end{equation*}
which means that an increase in the number of workers harms the ethical quality of workers when the average technical productivity of workers is already high.

Eq. (\ref{eq:quality}) incorporates, in an explicit way, two concepts in the theory of people risk management. First, the basic fraud model based on the Cressey's fraud triangle \citep[see][]{cressey1953other} asserts that fraud has three elements: Motivation or pressure to commit fraud, the opportunity to commit fraud, and the rationalization or justification that a fraudster makes to him or herself to commit fraud. An employee, often in dire financial straits, using its insider information about the firm's control system, redirects funds to other sources.

The insider information about the control environment is possible if the employee possesses knowledge about many processes in the bank. In this study, this knowledge is approximated by the technical productivity of workers in firm $i$ at time $\tau$: $a_{i,\tau}$

The second key fraud theory concept embedded in Eq. (\ref{eq:quality}) refers to interactions in the firm. In the words of \citet{blacker2015people}: ``Inappropriate interactions between individuals inside and outside of the firm give rise to People Risk'' (p. 121). Blacker and McConnell underscore the qualitative nature of employee interactions. In this study, it is argued that the qualitative level of interactions (inappropriate or bad) increases with the quantitative number of interactions that should be proportional to the number of employees scaled per branch $e_{i,t}$ at bank $i$ during period $t$.

Therefore, Eq. (\ref{eq:quality}) shows that the ethical quality of employees (inverse of the propensity to commit fraud) falls when both opportunities for fraud and the number of inappropriate interaction rise as suggested by the theory of people risk.

After losses $\{l_{i,\tau}\}$ for the set of banks $i=\{1, \ldots, N\}$ at high frequencies $\tau = \{1,\ldots,T\}$ are generated by the stochastic dynamic system given by Eqs. (\ref{eq:ramp}) to (\ref{eq:quality}), the loss data have to be recorded and submitted to the pooled database. 

Of note is that data recorded to build the loss datasets are not the same as the original loss data $\{l_{i,\tau}\}$ for a number of reasons. For example, the existence of recording thresholds indicates that only losses greater than a threshold level $l_i^{min}$ are submitted to the dataset. Moreover, when a loss event occurs, banks do not necessarily know the exact loss amount incurred. There is a natural lag between occurrence of an event that involves loss and knowledge of the severity of the event. The lag depends on the specific nature of the event. For the purposes of this research, it is assumed that the severity of the event is known at the same time as its occurrence but that the knowledge is imperfect and subject to measurement errors. Therefore, the observed dataset process implies
\begin{equation}
  \hat{l}_{i,\tau} = l_{i,\tau} + \eta_{i,\tau}
\end{equation}
where $\hat{l}_{i,\tau}$ is the observed loss severity, $l_{i,\tau}$ is the true unobserved loss severity (underlying loss) and $\eta_{i,\tau}$ is an unbiased, white noise measurement error distributed normally $\eta_{i,\tau} \sim N(0, \sigma^2_{\eta i})$. The measurement errors are independent over time and across banks, but heterogeneity across banks is allowed. The losses submitted to the pooled dataset and used for internal purposes are described by
\begin{equation}
  D_i = \{\hat{l}_{i,\tau}  |  \hat{l}_{i,\tau} > l_i^{min}\} \hspace{1cm}   \text{for } \tau = 1,\cdots, T
\end{equation}

In the data collection process, both $\sigma^2_{\eta i}$ and $l_i^{min}$ are assumed to be exogenous. In fact, the value of $l_i^{min}$ determined for all member banks of the ORX association is \euro 20,000.


\subsection{Model calibration} \label{subsec:calib}
The parameters that specify the stochastic dynamic operational loss model described above are not free. It is necessary to restrict the parameters to specific values. Standard estimation techniques such as linear regression cannot be conducted because there is no hard data and the model exhibits unobservable variables like control and quality of workers. This implies that the model parameters need to be calibrated. 
The operational loss simulation process is conditional on the value of these parameters.

Some parameters are specific to banks (idiosyncratic), so they have the subscript $i$ in their notation. Other parameters are common to all banks. The procedure for calibration of all parameters, general and specific, is described below. There are $13$ idiosyncratic parameters per bank (see Table \ref{tab:param1}). Given that there are up to $52$ banks, it would be necessary to pin-down about $672$ idiosyncratic parameters in total. Given the vast number of parameters to be calibrated, a very simple shrinkage procedure is introduced to reduce the number of parameters to be calibrated based on the available information, the period of analysis, and the specific banks under study.
\begin{table}
 \centering
  \caption{Parameters of the Dynamic Model for Operational Losses}
  \label{tab:param1}
\scalebox{.9}{
  \begin{tabular}{lll}
  \toprule
    Param.      	&Equation	            &Definition \\
    \midrule
    $\alpha_{i,0}$	&Internal fraud losses	&Overall scale of losses \\
    $\alpha_{i,0}$	&Internal fraud losses	&Constant within ramp function \\
    $\alpha_{i,c}$	&Internal fraud losses	&Impact of controls on losses   \\
    $\alpha_{i,y}$	&Internal fraud losses	&Impact of gross operating income on losses \\
    $\alpha_{i,q}$	&Internal fraud losses	&Impact of quality of workers on losses \\
    $\rho_i$	    &Internal fraud losses	&Autocorrelation of operational loss shocks \\
    $\beta_{0,i}$	&Internal fraud losses	&Constant in conditional variance of operational loss shocks \\
    $\beta_1$	    &Internal fraud losses	&Influence of past shocks on conditional variance of operational\\
                    &                       & loss shocks \\
    $\beta_2$	    &Internal fraud losses	&Influence of past variance on conditional variance of operational\\
                    &                       & loss shocks \\
    $\rho_c$	    &Loss control	        &Weight of new conditions to affect current controls \\
    $c_i^\ast$	    &Loss control	        &Control level associated to desired operational loss ratio \\
    $\gamma_i$	    &Loss control	        & Controls de sensitivity of control to the loss ratio gap from\\
                    &                       & the desired ratio \\
    $\lambda_i$	    & Loss control	        &Desired loss ratio \\
    $\rho_q$	    &Ethical quality	    &Weight of new conditions to affect current ethical quality levels \\
    $\delta_i$	    &Ethical quality	    &Determines the sign of impacts from factors \\
    $\bar{Q}$	    &Ethical quality	    &Level of average ethical quality across banks \\
    $\bar{A}$       &Ethical quality	    &Average labor productivity across banks \\
    $\bar{E}$       &Ethical quality	    & Average number of employees by branch across banks \\
    $\sigma^2_{\eta i}$	&Measurement error	    &Variance of measurement errors \\
    \bottomrule
  \end{tabular}}
\end{table}
In essence, the shrinkage procedure used in this study takes into account the idiosyncratic data collected for each bank. These data proxy the degree of riskiness and heterogeneity of each bank and are used to map the heterogeneous values of the model parameters.

The starting point for model calibration is the operational loss summary report presented in the ORX database \citep{orx2012} for the period 2006-2010. In this report, there are $4,357$ internal fraud loss events recorded in the retail-banking segment; the gross amount of losses reached \euro 880 million.

Table \ref{tab:orxbanks} in Appendix \ref{app:1} shows the banks that reported losses to the ORX data exchange during the period 2006-2010. New members entered the association, and some members quit due to bankruptcies, mergers, or acquisitions. For example, Wachovia was a member of the association until acquired by Wells Fargo in 2008.

For each of the $N$ banks and years under analysis, a set of variables categorized as key risk indicators are gathered. These variables condition the occurrence of losses in the model or serve as useful devices to calibrate idiosyncratic parameters. The set of conditioning variables is described in Table \ref{tab:param2}. All the variables indicate risk exposure such as the number of employees or the size of retail loans. These variables are useful devices to apply the shrinkage procedure because they discriminate among banks. For example, according to Cressey's fraud triangle explained before, banks that have higher employees per branch relative to the mean among banks might be deemed riskier than those that have lower employees per branch. Therefore, the dispersion of employees per branch across banks is useful to calibrate parameters across banks.
\begin{table}
\centering
  \caption{Observable Variables that Condition the Simulation of Losses in Each Bank}
  \label{tab:param2}
  \begin{tabular}{lll}
  \toprule
   Nomenclature	& Description	& Type \\
   \midrule
    $e_{i,t}$	& Number of employees	            &Idiosyncratic\\
    $b_{i,t}$	& Number of branches and offices	&Idiosyncratic\\
    $a_{i,t}$	& Retail assets (millions of Euros)	&Idiosyncratic\\
    $y_{i,t}$	& Retail loans (millions of Euros)	&Idiosyncratic\\
    $m_{i,t}$	& Proxy for operational risk management awareness	&Idiosyncratic\\
    $h_{i,t}$	& Proxy for human resource awareness	&Idiosyncratic\\
    \bottomrule
  \end{tabular}
\end{table}

In contrast to the model specification in the preceding subsection, the observed variables are indexed by time ($t$), where $t$ stands for end-of year variables. Idiosyncratic variables for the years $2006$ through $2010$ were obtained from annual reports that member banks published on their Web sites.\footnote{In the few cases when reports were not available, we got information from SEC filings.} The values of interest were extracted from the descriptive information, balance sheets, and income statements contained in the aforementioned reports. These reports are publicly available as part of the information disclosure by banks directed to investors. The financial statements in these reports are compatible with sound regulatory and accounting practices and, on the majority of cases, they accord to GAAP.

\begin{figure}[!ht]
\centering
    \caption{Number of branches and employees for banks in the ORX dataset} \label{fig:branches}
    \includegraphics[scale=0.65]{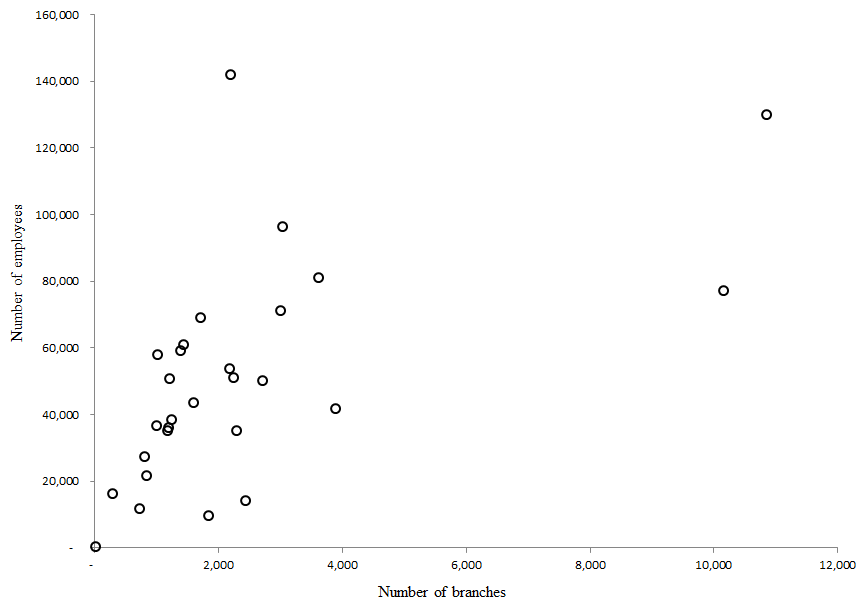}\\
    \footnotesize{Note: Information extracted from bank's annual reports as of December 2006}
\end{figure}

An example of the information recovered from these annual reports is given in Fig. (\ref{fig:branches}). The figure shows that the size of banks in the ORX dataset is heterogeneous. Each dot refers to a specific bank. The number of employees ranges from 10 to about 140 thousands, while the number of branches varies from 300 to about 10,000. In addition, both the number of employees and number of branches show some degree of correlation.

\begin{figure}[!ht]
\centering
    \caption{Word counts of the expression ``operational risk'' as percentage of page counts in each 2010 annual report.}
    \label{fig:textual}
    \includegraphics[scale=0.3]{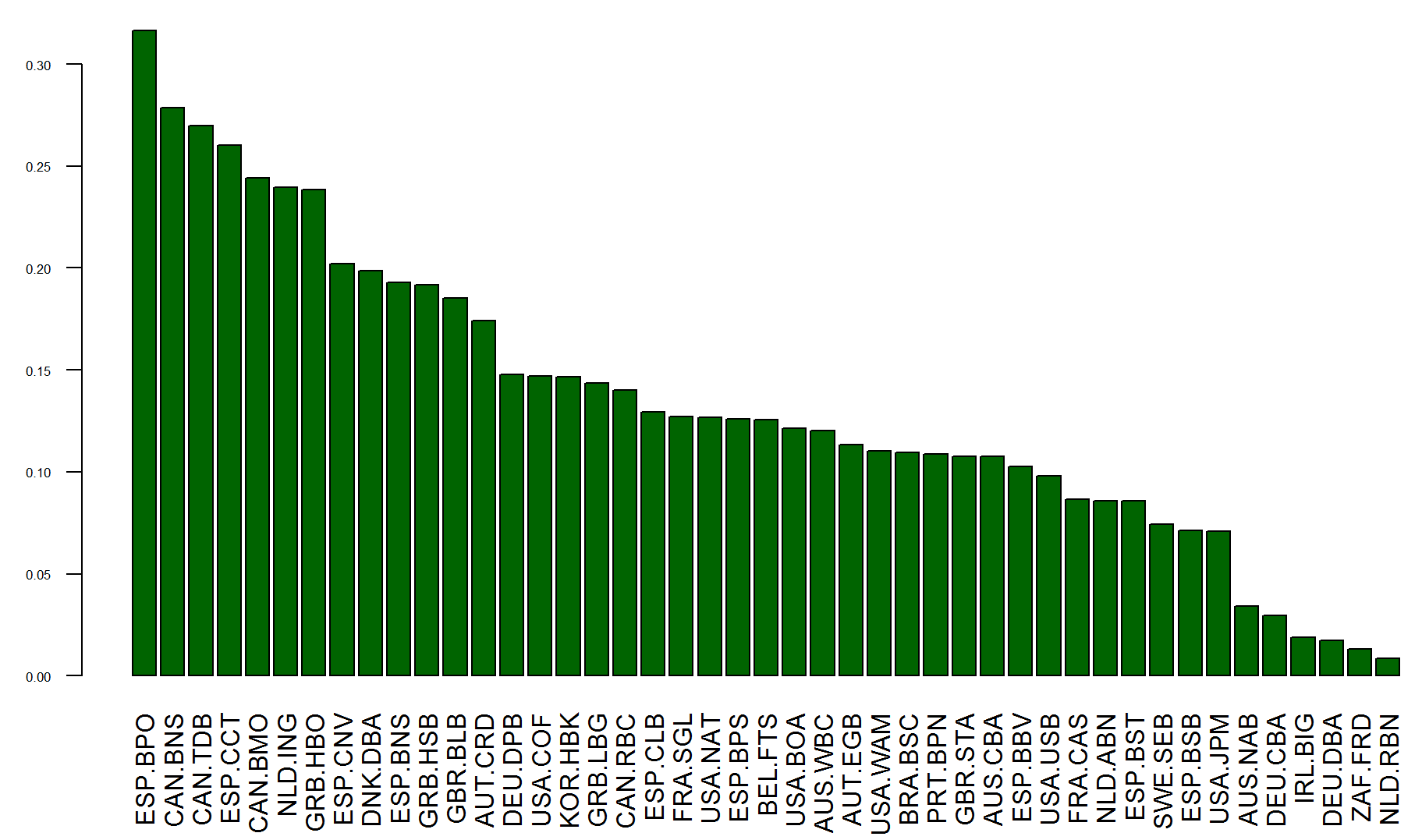}\\
    \footnotesize{Note: Information extracted from bank's annual reports as of December 2010}
\end{figure}

In addition to the objective information included in the annual reports, proxy variables related to operational risk controls and the quality of human resources at each bank were constructed. Thus, variable $m_{i,t}$ measures operational risk management awareness implicit in the information shared with the public. This awareness proxy was obtained by means of textual analysis of the published annual reports; for example, the number of times a word or a phrase occurred within each report divided by the number of pages. An example of this type of textual information is given by Fig. (\ref{fig:textual}), which depicts the ratio of word counts of the expression ``operational risk'' as a percentage of the total number of pages. This variable could arguably reflect the extent of awareness of each bank toward operational risk management.\footnote{The idea of extracting information from textual sources is not new in finance \citep{kearney2014textual}.}

Other textual expressions that reflect operational risk awareness are analyzed, for example, the use of the acronym AMA. 

The variable $h_{i,t}$ is intended to measure the awareness of banks about human resources. The annual reports also contain information about policies geared to improve the management and quality of human resources. So, human resource awareness could be obtained from textual analysis by extracting word counts of expressions such as ``employees'' or ``human resources''. The assumption is that these indicators reflect the quality of the workforce and are related inversely to the occurrence of internal fraud.

All the extracted information from banks' disclosures reflects the state of banks at calendar year-ends. Therefore, in order for these variables to be entered into the simulation model, it is necessary to perform simple linear interpolations to complete data for all moments of time $\tau$ between any consecutive years $t$ and $t+1$. It is assumed that time $\tau$ will refer to business days within years. The existence of holidays was excluded in these calculations.

Five parameters affect the outbreak of operational losses in Eq. (\ref{eq:ramp}). All these parameters are idiosyncratic; therefore, it was necessary to devise a way to calibrate all of them in a simple form. Let $\alpha_{i,j}$ be a parameter in Eq. (\ref{eq:ramp}) for $i =\{1,\cdots,N\}$, and $j = \{0,1,c,y,q\}$. Then for each $j$, there exists a mean parameter value taken from the cross section of banks. $\bar{\alpha}_j = \sum_{i=1}^N \alpha_{i,j}$. By pinning down the value of $\bar{\alpha}_j$, it is possible to pin down the idiosyncratic parameters of interest $\alpha_{i,j}$.

\begin{figure}[!ht]
\centering
    \caption{Histogram of the number of employees per branch across banks in the ORX database as of December 2006.} \label{fig:ehisto}
    \includegraphics[scale=0.8]{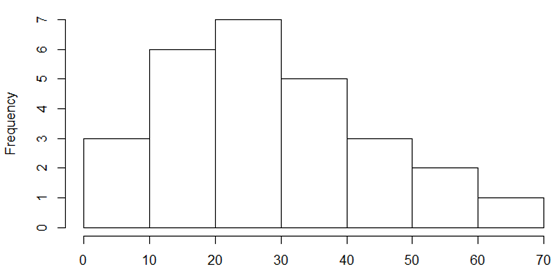}\\
    \footnotesize{Note: Information extracted from bank's annual reports as of December 2006}
\end{figure}

To complete the process, it is necessary to work with risk exposure indicators calculated from the data shown in Table \ref{tab:param2}. Let these indicators be defined by $x_{i,t}$, for example, the ratio of employees per branch or the technical quality of workers (labor productivity) measured as the ratio of total retail loans to the number of employees. These measures can be calculated for each financial institution and for each calendar year in the sample. Fig. (\ref{fig:ehisto}) depicts a histogram of the number of employees per branch at the end of 2006 in all banks belonging to the dataset by the end of that year.

Next, let  $\bar{x} = \frac{1}{NT}\sum_{t=1}^T\sum_{i=1}^N x_{i,t}$  be the overall average risk indicator and $\bar{x}_i = \frac{1}{T}\sum_{t=1}^T x_{i,t}$ be the bank $i$ average. If there are grounds to postulate direct proportionality between the coefficient $\alpha_{i,j}$ and the indicator $\bar{x}_i$, then each parameter can be pinned down according to
\begin{equation}
  \alpha_{i,j} = \big( \frac{\bar{x}_i}{\bar{x}}\big)\bar{\alpha}_j\hspace{2cm} \text{for } i= 1, \cdots, N   \label{eq:calib1}
\end{equation}

For example, for parameters $\alpha_{i,0}$, $\alpha_{i,1}$, and $\alpha_{i,y}$, the choice of $x_{i,t} = \frac{e_{i,t}}{b_{i,t}}$  is a reasonable option. This means that loss event sensitivities are correlated with the ratio of employees to banks. In this case, only the parameter $\bar{\alpha}_j$ need be calibrated. Once its value is determined, Eq. (\ref{eq:calib1}) fixes the distribution of $\alpha_{i,j}$ parameters across banks.

The parameters $\alpha_{i,c}$ and $\alpha_{i,q}$ are likely to be inversely proportional to the ratio $x_{i,t}=\frac{e_{i,t}}{b_{i,t}}$. For example, controls are more effective when there are fewer people working at branches.\\
The adjustment can be made according to
\begin{equation}
  \alpha_{j,i} = \big( \frac{\bar{x}}{\bar{x}_i}\big)\bar{\alpha}_j\hspace{2cm} \text{for } i= 1, \cdots, N
\end{equation}
Parameters $\rho_i$ and $\beta_{0,i}$ can be adjusted in the same fashion. In the case of the autocorrelation of shocks $\rho_i$, it is necessary to set bounds $\rho_{min}>0$ and $\rho_{max}<1$, such that the resulting operation $\rho_i' = \frac{\bar{x}_i}{\bar{x}}\rho$ can be further modified to become bounded within the range $[\rho_{min},\rho_{max} ]$. To do so, it is first necessary to calculate $\rho'_{min}$ and $\rho'_{max}$ with the parameters obtained given $\rho$ and then to apply
\begin{equation}
  \rho_i = \rho_{min} + \bigg(\frac{\rho_{max}-\rho_{min}}{\rho'_{max}-\rho'_{min}} \bigg) (\rho'_i - \rho'_{min}) \hspace{2cm} \text{for } i= 1, \cdots, N
\end{equation}
The parameter $\beta_{0,i}$ controls for the unconditional variance of operational loss shocks in each bank, as shown in Eq. (\ref{eq:shock}). The dispersion of this parameter across banks can also apply the principle underlying Eq. (\ref{eq:calib1}).

Three idiosyncratic parameters affect the level of controls $(c_i^\ast,\lambda_i,\gamma_i)$. The first parameter measures the long-run value of control levels. Current control levels may be stricter or easier than this long-run benchmark, which also needs to be on the range $[0,1]$. To calibrate the dispersion of this parameter, the indicator $m_{i,t}$ that measures operational risk management awareness can be used in the same fashion as the calibration of the dispersion of parameter $\rho_i$.

Parameter $\lambda_i$ reflects the operational loss level as a ratio of operating income that banks are ready to accept. Operational loss ratios larger than this benchmark $\lambda_i$ prompt banks to increase their controls. Calibration of this parameter for each bank is problematic because the available information does not provide reasonable proxies for this ratio. Therefore, this paper assumes that this ratio is similar across all banks. Its level is determined by the median 2008 ratio of cumulative operational losses to gross operating income provided by \citet{benyon2008top}.

Parameter $\gamma_i$ is a feedback adjusting parameter. From Eq. (\ref{eq:control}), it is easy to note that for controls to tighten whenever the operational loss ratio increases, the parameter $\gamma_i$ has to be negative. The greater $\gamma_i$ is in absolute value, the greater the impact on control levels. Again, it is reasonable to assume that the absolute value of $\gamma_i$ is directly proportional to the level of operational risk awareness $m_{i,t}$ and thus, the calibration of the dispersion in $\gamma_i$ will apply the same steps outlined above.

In terms of the ethical quality of human resources in a bank, the parameter $\delta_i$ measures how sensitive each bank's workforce quality is to the bank's size and labor productivity. Eq. (\ref{eq:quality}) assumes that size and labor productivity may be detrimental to workers ethical quality. Therefore, parameter $\delta_i$ is positive, and the higher it is, the more sensitive ethical quality becomes. The sensitivity may be related inversely to the human resource awareness proxy $h_{i,t}$ extracted from the data. Human resource awareness is related to how important is the workforce in terms of well-being, compensations, and on-the-job training, for example.

\begin{figure}[!ht]
\centering
    \caption{Histogram of aggregate 12-month cumulative operational losses for October 2008 in banks in the ORX database.} \label{fig:benyonhisto}
    \includegraphics[scale=0.6]{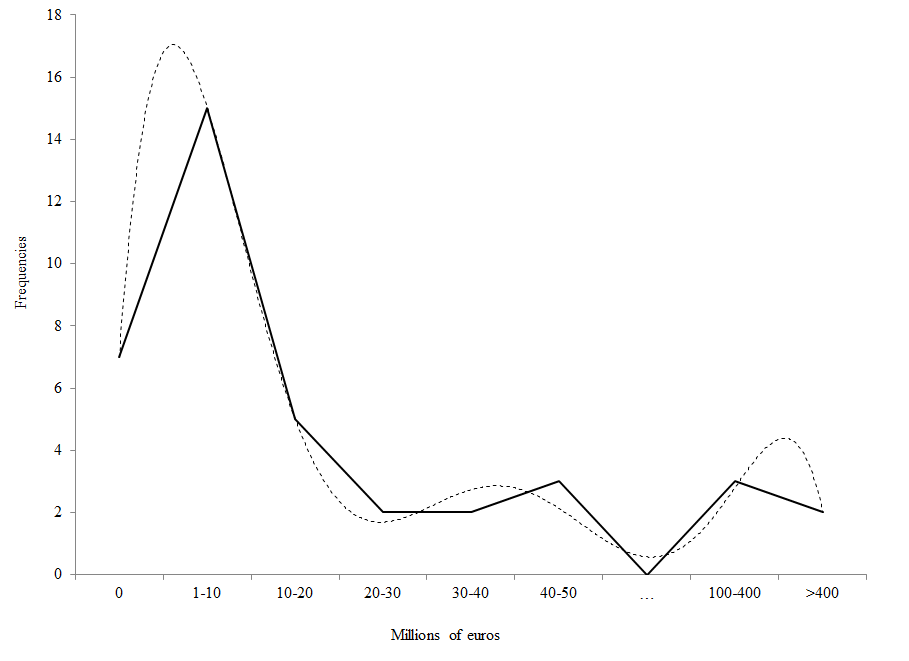}\\
    \footnotesize{Note: For each bank, average percentages are reported. Ranges of losses are expressed in millions of Euros. The continuous line is the histogram; the dashed line is a polynomial smoothing of the original histogram. Data were obtained from \citet{benyon2008top}}
\end{figure}

The last idiosyncratic parameter calibration that needs a shrinkage procedure is the variance $\sigma_{\eta i}^2$ of the measurement error in recording the severity of operational losses. It is reasonable to assume that higher severity levels are associated with higher measurement error variances.  To reflect this feature, the study used the aggregate operational loss figures reported in \citet{benyon2008top}. The losses reported by Benyon refer to the aggregate of all types of operational losses, not just retail banking.  A summary of this data is depicted in Fig. (\ref{fig:benyonhisto}), which shows the existence of an extreme asymmetry of operational loss severities; in fact, two important modes appear for losses less than \euro 10 million and for losses larger than \euro 100 million. It is assumed that banks facing large loss severities are likely to have large variances in their measurement errors when they record operational losses. Therefore, the dispersion in $\sigma_{\eta i}^2$ will be calibrated by the dispersion of loss severities documented in \citet{benyon2008top}.

After the shrinkage procedure is executed, the space of parameters to calibrate shrinks to 20. The mean value $\bar{\alpha}_j$ of the parameters in the operational loss (Eq. \ref{eq:ramp}) was calibrated to generate loss severities compatible with the 2012 summary of the database for retail-banking losses due to internal fraud \citep{orx2012}. The mean value of the parameter $\bar{\beta}_0$ was set to pin down the frequency of losses documented in the ORX summary. Parameters $\bar{\rho}$, $\beta_1$ and $\beta_2$ determined the clustering pattern of operational loss shocks. According to \citep{chernobai2008dynamics}, internal fraud losses exhibit low clustering as opposed to other type of losses. Hence, the calibration assigns relatively low values to these parameters in the range 0.01 and 0.1.

Two parameters measure speed of response. First, $\rho_c$ measures how quickly internal controls are implemented to achieve a new level after new conditions arise and are expected to persist. Second $\rho_q$ measures how fast the average ethical workforce achieves a new level when conditioning factors change and are expected to last.

Regarding $\rho_c$, in broad terms, control levels do not change from one day to another, and some changes necessary to implement control adjustments may require budgeting, planning, and extra human resources. In a given year, the average worst scenario would be to wait half a year to implement full changes. Thus, if the number of business days in a year is about 260 and the number of working days to implement a new long-run control level is 130, approximately 65 days are necessary for implementing half the changes. Due to the auto-correlated nature of controls in Eq. (\ref{eq:control}) and a half-life of $\tau' = 65$ days, the parameter $\rho_c$ is then set to the value $\rho_c = 1 -(\frac{1}{2})^{1/\tau'} \approx 0,011$. Changes in the average level of ethical quality of the workforce may take even more time, and therefore, the parameter $\rho_q$ would have to be lower than the benchmark of 0,011.

Two parameters need to be determined in the control equation (Eq. \ref{eq:control}), namely the mean value of long run control levels $\bar{c}^\ast$ and the mean sensitivity of control to loss ratio $\bar{\gamma}$. In the ethical quality equation (Eq. \ref{eq:quality}), there are four parameters to be set; two of them being the average labor productivity $\bar{A}$ across banks and the average number of employees per branch across banks $\bar{E}$ , which are both readily estimable from the data.

All the remaining parameters grouped in the vector $(\bar{c}^\ast,\bar{\gamma},\bar{\delta},\bar{Q},\bar{\sigma}_\eta^2)$ are set freely with the hope that given their benchmark values, the simulated loss severities and frequencies are associated with external observed factors such as macroeconomic and institutional variables.
\section{The data}\label{sec:data}
Most but not all the data necessary to perform the analysis belongs to the ORX data exchange. However, this dataset is proprietary. Instead, the data gathered for the analysis in this paper relies entirely on public information. All banks publish their annual reports and financial statements each year and sometimes more often. These reports do not contain information about operational losses but contain most of the idiosyncratic data needed for model calibration. Table \ref{tab:reportvars} summarizes the type of data collected from each of the reports or financial statements.
\begin{table}
\centering
  \caption{Data Gathered from Public Sources about Banks in the ORX Exchange}
  \label{tab:reportvars}
  \begin{tabular}{lll}
  \toprule
    Key	      & Concept	                    & Measure\\
    \midrule
    No	      & Number of bank	            &Index\\
    bname     &	Bank name	                &Index\\
    country   &	Bank headquarters' country  &	Index\\
    code	  & Country and Bank Code	    &Index\\
    year	  & Year	                    &Index\\
    ccy	      & Report's Currency Code	    &Text\\
    branches  & Number of branches and offices (Retail)	&Count\\
    staff	  & Number of staff (total)	    &Count\\
    staff\_r  & Number of staff (Retail)    &	Count\\
    loans	  & Total loans to customers    &	Currency millions\\
    loans\_r  & Total loans to customers in retail banking	&Currency millions\\
    assets	  & Consolidated assets         &	Currency millions\\
    assets\_r & Assets in retail banking	&Currency millions\\
    tier1	  & Tier 1 capital	            &Percent\\
    nic	      & Net interest income (total)	&Currency millions\\
    nic\_r	  & Net interest income (retail)&Currency millions\\
\bottomrule
  \end{tabular}
\end{table}

\begin{figure}[!ht]
\centering
    \caption{Scatter plot of bank data per year.} \label{fig:orxdata}
    \includegraphics[scale=0.75]{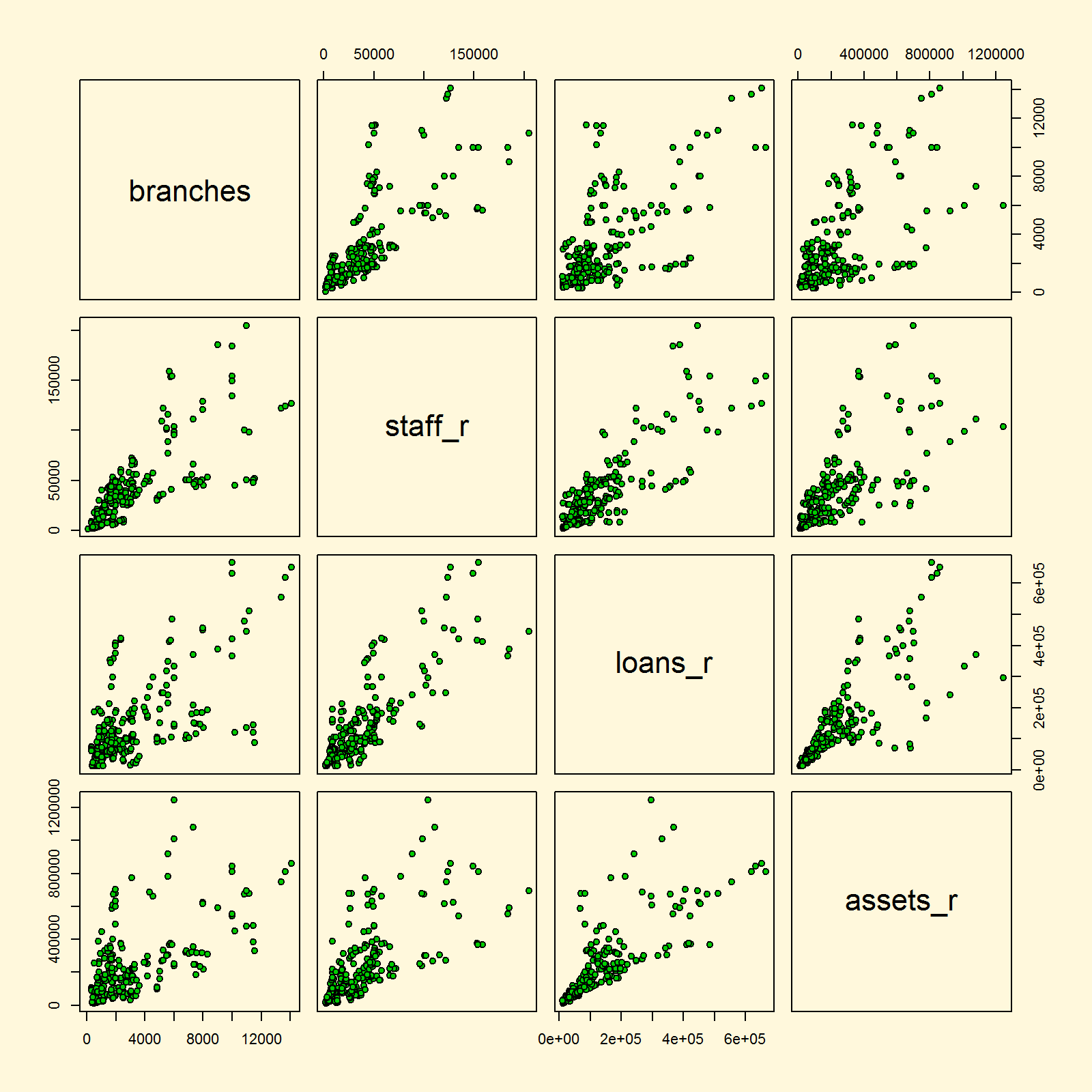}\\
\end{figure}

Fig. (\ref{fig:orxdata}) provides a brief description of the dataset. The figure shows pairs of scatterplots between the numbers of branches, the number of staff related to retail banking operations, the level of retail loans, and the value of retail assets. All currency amounts were converted to millions of Euros. Data comprised five years for all 52 banks considered.

All variables considered in Fig. (\ref{fig:orxdata}) are indicators of the scale of operations in the retail banking segments of each bank. This explains the remarkable positive correlation between 0.62 and 0.82. These scale indicators belong to the set of risk indicators that likely induce the appearance and severity of internal fraud losses.

The calibration procedure needs more specific risk indicators. Hence, the analysis relies on other forms of risk indicators that could be collected from annual reports. Textual content is useful to calibrate sensitivity parameters as shown in Section (\ref{sec:model}). Table (\ref{tab:textual}) shows the textual context variables extracted from the annual reports. The variables refer to the number of instances a descriptive key word or phrase appears within the entire texts; also, the total number of report pages is recorded in order to calculate the ratio of instances to number of pages. These ratios give an indication about the relative importance of a key word that banks use in their public reports.
\begin{table}[!ht]
\centering
  \caption{Variables Contained in the Textual Database}
  \label{tab:textual}
  \scalebox{0.9}{
  \begin{tabular}{ll}
    \toprule
    Key	    &Concept\\
    \midrule
    nbank	&Number of bank in database\\
    bname	&Bank name\\
    country	&Country of bank headquarters\\
    code	&Bank Code = country code.bank\\
    year	&Year\\
    orisk	&``Operational risk'' term frequency in annual reports\\
    risk	&``Risk'' term frequency in annual reports\\
    rman	&``Risk management'' term frequency in annual report\\
    ama	    &``AMA'' (Advanced Measurement Approach) term frequency in annual reports\\
    hres	&``Human resource'' term frequency in annual reports\\
    emp	    &``employee'' term frequency in annual reports\\
    Col	    &``colleague'' term frequency in annual reports\\
    workers	&Sum of ``employee'' and ``colleague'' term frequencies\\
    npag	&Number of pages in the Annual Report\\
    \bottomrule
  \end{tabular}}
\end{table}

\begin{figure}[!ht]
\centering
    \caption{Scatter plot of textual context variables.} \label{fig:textual2}
    \begin{tabular}{cc}
      Panel A                                          & Panel B \\
     \includegraphics[scale=0.35]{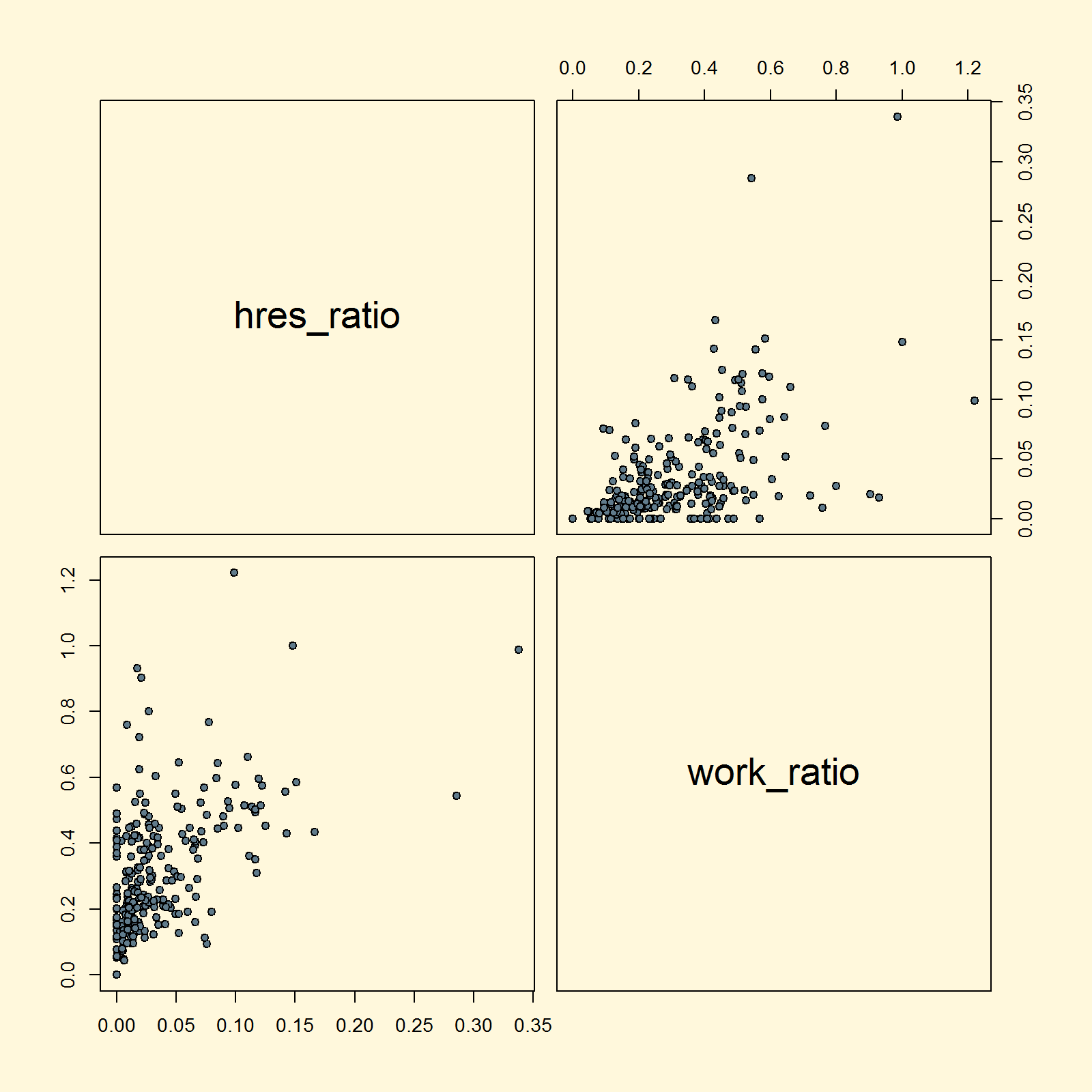} &  \includegraphics[scale=0.35]{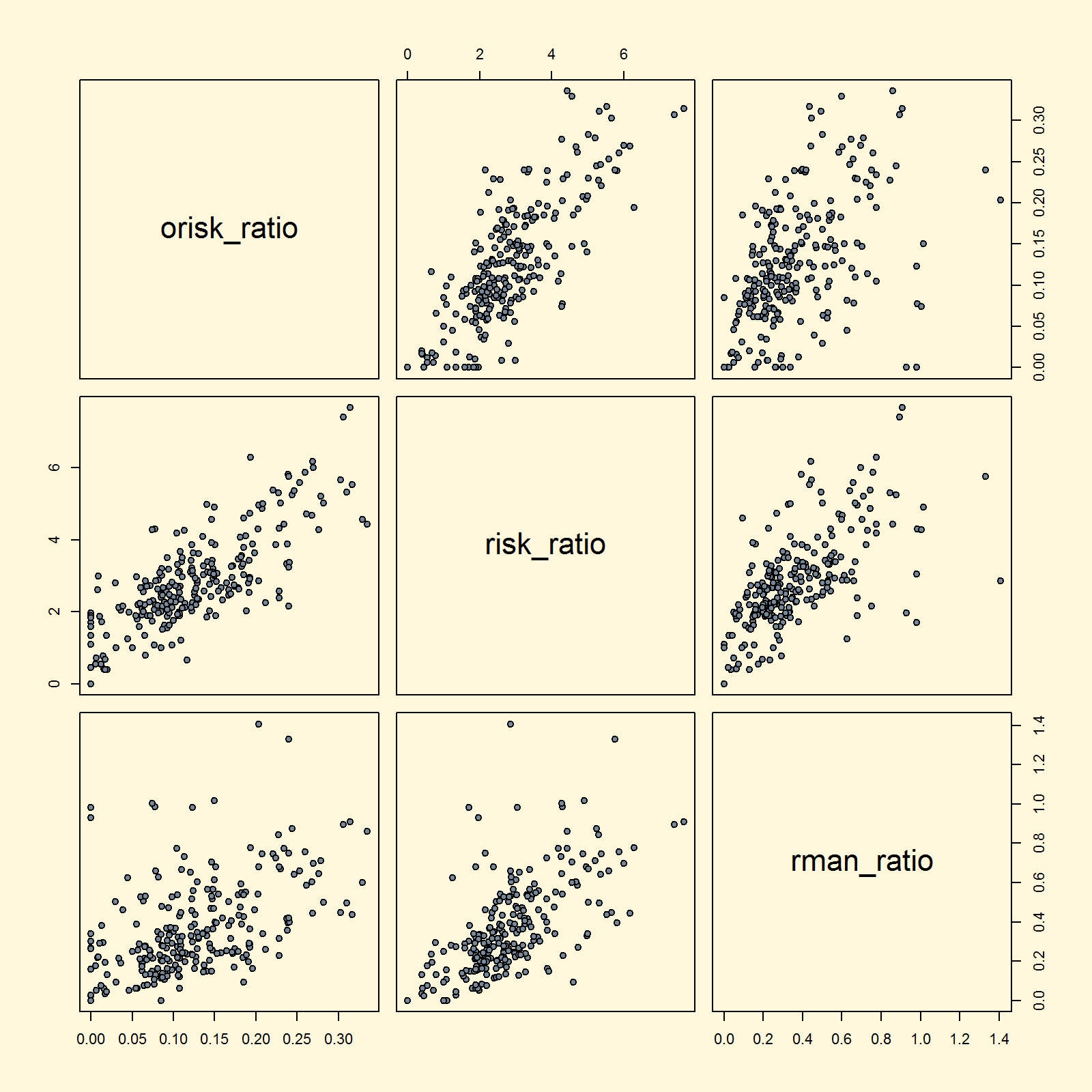}\\
    \end{tabular}
\end{figure}

Both panels in Fig. (\ref{fig:textual2}) show scatter plots of textual variables. Panel A contains variables related to human resource management. It shows scatterplots of ``human resource'' paired with the sum of ``employee'' and ``colleague'' instances in annual report texts as a proportion of total number of pages in each report.  Panel B contains variables related to risk management. It shows the paired scatter plot of the triplet of risk, risk management, and AMA instances as a proportion to total pages. All plots show positive correlations. 

In terms of global variables that affect all banks or groups of banks, the analysis incorporates variables such as GDP growth rates in the countries to which banks belong. The dataset also contains a number of variables that could affect the outbreak of losses due to internal fraud such as the rule of law in a country or its corruption perception index.
\begin{table}[!ht]
\centering
  \caption{Variables Contained in the Macroeconomic Database}
  \label{tab:macro}
  \begin{tabular}{ll}
  \toprule
    Key	               &Indicator name\\
    \midrule
    country\_name	   &Country Name\\
    country\_code	   &Country Code\\
    Year	           &Year 2006-2010\\
    gdp\_growth	       &GDP growth (annual \%)\\
    crisis	           &Financial crisis dummy variable for 2007 and 2008\\
    gover\_effective   &Government Effectiveness\\
    reg\_quality	   &Regulatory Quality\\
    rule\_law	       &Rule of Law\\
    cont\_corrup	   &Control of Corruption\\
    cpi         	   &Corruption Perceptions Index (CPI) score (2006-2010)\\
    \bottomrule
  \end{tabular}
\end{table}

\begin{figure}[!ht]
\centering
    \caption{Corruption perception index (CPI) for countries where banks have their main headquarters.} \label{fig:cpi}
     \includegraphics[scale=1]{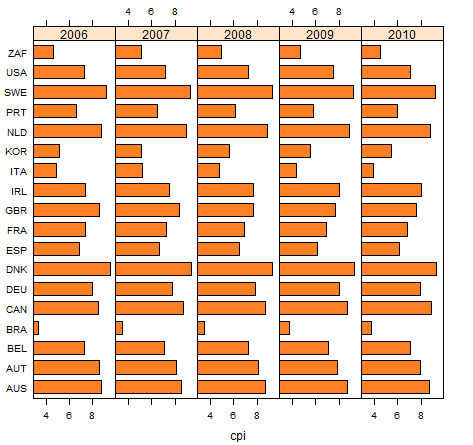}\\
     \footnotesize{Note: The lower the index, the more a country is perceived to be corrupt. The data were based on figures released by Transparency International.}
\end{figure}

Fig. (\ref{fig:cpi}) shows the corruption perception index (CPI) for each of the relevant countries during the years of analysis. In the data, Brazil and Italy are shown to have higher corruption perceptions while countries like Denmark, Sweden, Netherlands, and Canada are seen to be less likely to be corrupt. The perception of corruption is possibly associated with real corruption levels, and the extent of corruption can affect the occurrence of fraud internal or external to banks because they are related to the cultural environment that rationalizes frauds according to Cressey's triangle.


A good measure of the fitness of the operational risk model to generate internal fraud losses is whether those losses are associated to the macro-risk indicators in the same fashion as documented in empirical research as shown for example in \citet{chernobai2011determinants}, \citet{moosa2011operational}, \citet{cope2012macroenvironmental}, \citet{abdymomunov2017us}, and \citet{stewart2016bank}.

\section{Results} \label{sec:result}
First, we report the results of the calibration procedure. Second, conditional on calibrated parameters, model simulations that capture the loss profile are performed, given the environment and conditions that banks faced during the period 2006-2010 (See Table \ref{tab:orxbanks}). Third, with the simulated data for each bank and their corresponding conditioning factors, regression results that show the link between macro variables and operational losses are presented.

\subsection{Calibration results}
Table \ref{tab:calib} shows the value of the general parameters that set the behavior of the equations in the operational loss model. The setting of these parameters applied the calibration procedure outlined in Section \ref{sec:model}. The target of the calibration is to allow the model to simulate aggregate losses as close to reality as is possible. The only reality check available was to mimic the mean frequency and severity of losses due to internal fraud in the retail segment across the banks belonging to the ORX data exchange for the period 2006-2010. Therefore, the calibration procedure uses an optimizing framework to pin down the mean parameters of the loss equation described in Table \ref{tab:calib}.
\begin{table}[!ht]
 \begin{threeparttable}
  \caption{Calibration of Parameters}
  \label{tab:calib}
  \begin{tabular}{llcl}
  \toprule
	                        &Definition	                                     &Value	     &Eq.\\
  \midrule
    $\bar{\alpha}_0$	    &Mean scale parameter	                         &    0.400	 &Loss outbreaks\\
    $\bar{\alpha}_1$	    &Mean constant within ramp function	             &   16.810	 &Loss outbreaks\\
    $\bar{\alpha}_c$	    &Mean impact of controls	                     & -275.291  &Loss outbreaks\\
    $\bar{\alpha}_y$	    &Mean impact of gross operating income	         &   -1.587	 &Loss outbreaks\\
    $\bar{\alpha}_q$	    &Mean impact of quality of workers on losses     &	  0.052  &Loss outbreaks\\
    $\bar{\rho}$	        &Mean autocorrelation of loss shocks	         &    0.70   &Loss shocks\\
    $\rho_{min}$	        &Lower threshold for loss shocks autocorrelation &	  0.50   &Loss shocks\\
    $\rho_{max}$	        &Upper threshold for loss shocks autocorrelation &	  0.90   &Loss shocks\\
    $\bar{\beta}_0$	        &Mean constant term	                             &    0.20	 &Shock variance\\
    $\beta_1$	            &Influence of past quadratic shocks	             &    0.01   &Shock variance\\
    $\beta_2$	            &Influence of past variance	                     &    0.70   &Shock variance\\
    $\rho_c$	            &Weight of new conditions to affect controls	 &    0.10   &Loss control\\
    $\bar{c}^\ast$	        &Mean long run value of control level	         &    0.5    &Loss control\\
    $\bar{c}_{min}$	        &Lower threshold for long run control level	     &    0.3	 &Loss control\\
    $\bar{c}_{max}$	        &Upper threshold for long run control level	     &    0.7	 &Loss control\\
    $\bar{\gamma}$	        &Sensitivity of controls to the losses	         &   -0.5	 &Loss control\\
    $\lambda$	            &Desired loss ratio	                             &    0.0003 &Loss control\\
    $\rho_q$	            &Sensitivity to recent ethical quality	         &    0.05	 &Ethical quality\\
    $\bar{\delta}$	        &Determines the sign of impacts from factors	 &    0.2	 &Ethical quality\\
    $\bar{Q}$	            &Level of average ethical quality across banks	 &    0.7	 &Ethical quality\\
    $\bar{\sigma}_\eta^2$   &Variance of measurement errors	                 &    0.012	 &Measurement error\\
    $l_i^{min}$	            &Threshold level for operational loss reporting	 &   20	     &Reporting\\
    \bottomrule
  \end{tabular}
    \begin{tablenotes}
      \footnotesize{
      \item Note: The shrinkage procedure uses the parameters denoted with overbars. An optimizing search procedure determines the $\alpha$ parameters.}
    \end{tablenotes}
  \end{threeparttable}
\end{table}

The optimizing framework hinges on minimizing the quadratic distance between the observed mean loss severity and the simulated mean loss severity. In addition, the optimization puts weight on the fact that almost all banks in the dataset must face losses. In reality, a bank with no operational losses in the pace of five years is rare. The parameters that affect banks in an idiosyncratic way are determined by the shrinkage procedure described in Subsection \ref{subsec:calib}. Figs. (\ref{fig:app1}) and (\ref{fig:app2}) in the appendix depict the distribution of these parameters across banks. The idiosyncratic parameters calibrated through this procedure therefore serve as a useful device to control for the heterogeneity observed in the banks in the ORX sample.

\subsection{Simulation results and analysis}
\begin{figure}[!ht]
\centering
    \caption{Summary of simulations and comparison to ORX report.} \label{fig:sim}
     \includegraphics[scale=.35]{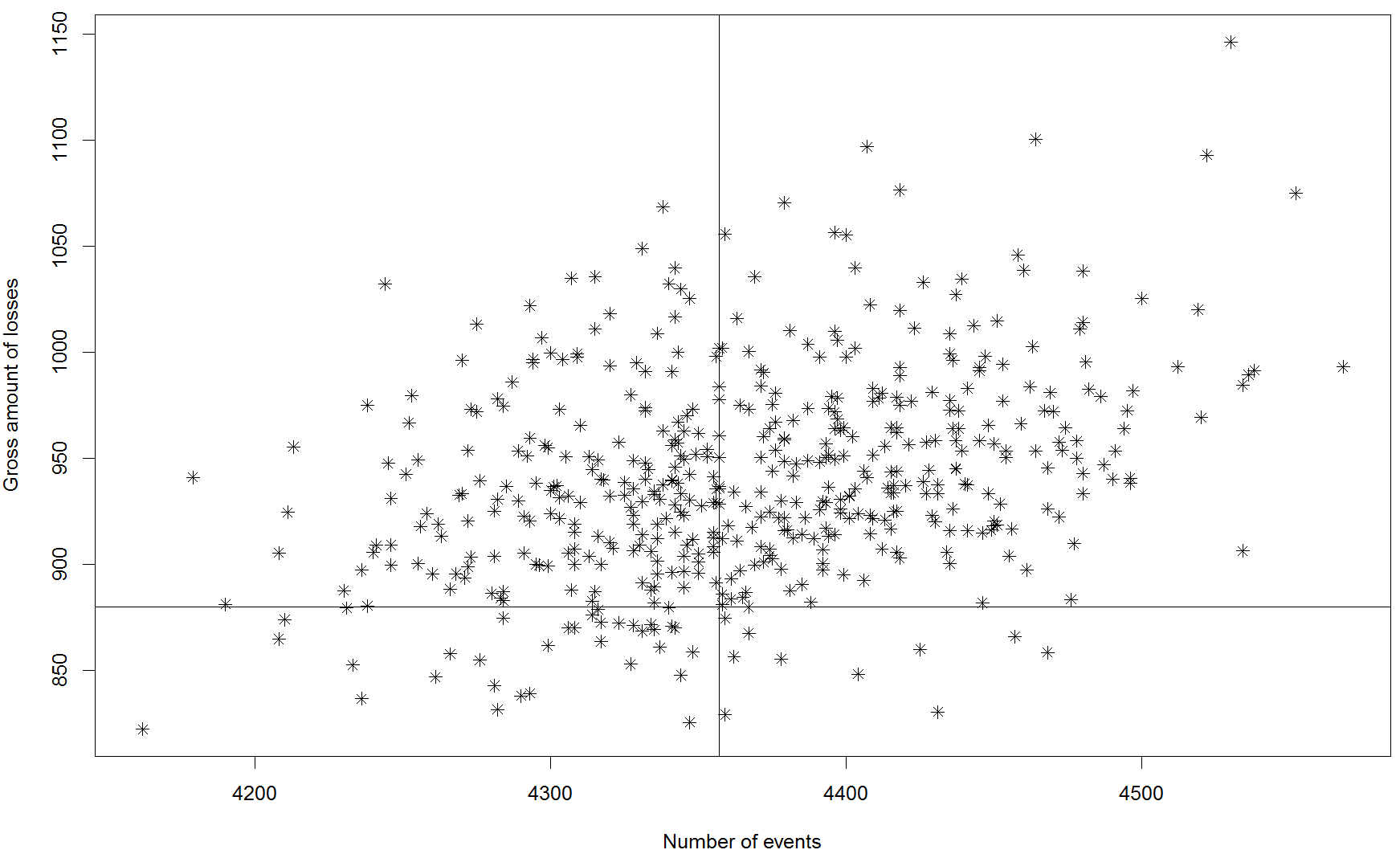}\\
\end{figure}

This study simulates 500 alternative histories of operational losses for the years 2006-2010 within the banks in the ORX database by drawing from the shocks in Eq. (\ref{eq:ramp}). The simulations consider the specific conditions banks confronted during the five-year period in terms of their own risk exposure and controls implemented. After each simulation, the gross amount of operational losses as well as the number of losses across banks were calculated. The 500 data points are drawn in Fig. (\ref{fig:sim}), where the straight lines mark the actual values reported in \citet{orx2012}.

Each point in Fig. (\ref{fig:sim}) summarizes a possible five-year history of data aggregated across each of the 52 banks. Each bank has 500 possible histories of operational losses conditional to the circumstances in effect during those five years. 

As a model validation step, this paper studies the association between simulated operational losses and covariates based on each bank and covariates that reflect the macroeconomic environment. Given the heterogeneity of operational loss data, this paper models the loss frequency distribution and the loss severity density in such a way that the location and scale parameters of these functions are affected by the set of covariates.

The paper uses the generalized additive models for location, scale, and shape (GAMLSS) as developed in \citet{stasinopoulos2007generalized} and used by \citet{ganegoda2013scaling} in the operational risk context. Let the simulated data be given by counts of loss outbreaks $\{\hat{n}_{i,T}\}$ for each bank $i$ and during years $T$ and sequences of loss severities (valued in Euros) $\{\hat{l}_{i,\tau}\}$ during the period under analysis. Severity observations and yearly count (frequencies) can be modeled using standard statistical distributions used in the operational risk literature.

The GAMLSS approach implies that covariates affect the location and scale parameters of these distributions. If the severities are $\hat{l}_{i,\tau}$, the GAMLSS method assumes that they are drawn from a density function $f(\hat{l}_{i,\tau};\Theta_i)$ conditional on $\Theta_i$.  In this paper, the conditioning parameter vector is given by $\Theta_i = (\tilde{\mu}_i, \tilde{\sigma}_i)'$, where $\tilde{\mu}_i$ stands for location and $\tilde{\sigma}_i$ stands for scale.  Both parameters are linked to covariates through link functions
\begin{equation*}
  g_1(\tilde{\mu}_i) = Z_1 \omega_1
\end{equation*}
and
\begin{equation*}
  g_1(\tilde{\sigma}_i) = Z_2 \omega_2
\end{equation*}	
where $Z_1$ and $Z_2$ are covariates that affect the distributional parameters and the $\omega 's$ are the corresponding sensitiveness coefficients.

The regressions comprise the mean and the scale parameters using a number of distributional assumptions and specifications about the behavior of the mean or the scale of the distributions as functions of the covariates. In contrast to simple OLS regressions\footnote{We performed but not reported standard OLS regressions.}, the GAMLSS method tackles the heteroscedastic nature of the data in a straightforward way.

\noindent \textbf{Severity regressions}

Table \ref{tab:lossreg} shows the results for the best performing model for loss severities in retail banking associated with internal fraud. This model is the truncated Weibull with mean and scale parameter.\footnote{Given that losses smaller than \euro 20,000 are not considered in the data, all loss data is truncated from below and therefore, truncated distribution functions needed to be estimated.}.

Results show that when a country to which a bank belongs to grows, the average size of losses increases. This result is similar to findings reported in \citet{povel2007booms} and \citet{stewart2016bank}. This positive impact has to do with the opportunistic behavior of workers stressed by \citet{blacker2015people}. Arguably, in general macroeconomic boom periods more fraud opportunities arise.

In addition, when a country is perceived as more corrupt (lower CPI), the average losses are higher. In the model section, our paper points out that the level of bad interactions inside and outside banks bring about a rationalization for fraud. The corruption perception of a country is a proxy for outside bad interactions. In this sense, if criminal or corrupted behavior of citizens is broadly accepted in a society, then workers find more rationale for stealing from banks.

The regressions also consider idiosyncratic variables. For example, higher levels of operational risk controls reduce loss severities, higher employees per branch increase loss severities and higher assets per employee reduce loss severities. These three effects are expected because they are embedded in the loss model calibration or specification.

Nevertheless, the remarkable finding is that neither the GDP growth nor the CPI were used to calibrate the loss model or as causal variables in the model specification, yet they significantly cause fraud losses. Furthermore, this finding conforms well with existing theory of how internal fraud losses occur.

\begin{table}[!ht]
\centering
\begin{threeparttable}
\caption{Estimates of the Regression in the Mean (Truncated Weibull)}
  \label{tab:lossreg}
  \begin{tabular}{lccccc}
  \toprule
    Regressors	             &Estimates	&Std. error	  &t-value  &p-value & \\
    \midrule
    Regression in the mean   &           &            &         &        & \\
    \midrule
    Intercept	            &-2.80	    &0.098	      & -28.43	&0.000	&***\\
    GDP growth	            & 0.01	    &0.006	      &   2.09	&0.037	&*\\
    CPI	                    &-0.05	    &0.012	      &  -4.06	&0.000	&***\\
    Control	                &-0.40	    &0.184	      &  -2.17	&0.030	&*\\
    Employees per branch	& 0.13	    &0.003	      &  51.57	&0.000	&***\\
    Assets per employee	    &-0.01	    &0.005	      &  -1.95	&0.051	&.\\
    \midrule
    Regression in the scale &           &             &         &       & \\
    \midrule
    Intercept	            &  0.733	&0.085	      &   8.61	&0.000	&***\\
    CPI	                    &-0.068	    &0.011	      &  -6.32	&0.000	&***\\
    Employees per branch	&-0.012	    &0.002	      &  -5.84	&0.000	&***\\
    GDP growth (a)	        & 0.002	    &0.005	      &   0.45	&0.651	&\\
    Control (a)	            &-0.015	    &0.153	      &  -0.10	&0.924	&\\
    Assets per employee (a)	& 0.009	    &0.005	      &   1.85	&0.065	&.\\
    \bottomrule
  \end{tabular}
  \begin{tablenotes}
      \footnotesize{
      \item Notes: Significance codes: $0 = ***$, $0.001=**$, $0.01=*$, $0.05=.$
      \item (a) Smoothing is performed with p-splines}
    \end{tablenotes}
    \end{threeparttable}
\end{table}
Table \ref{tab:lossreg} also shows the results for the scale parameter of the truncated Weibull distribution. The results suggest that a higher corruption perception (lower index) implies higher fraud loss volatility, but more employees per branch diminish the fraud loss volatility. These results are not straightforward to justify in terms of theory but provides an interesting starting point for further research. For now, this is beyond the scope of this paper.

Table \ref{tab:lossreg} reports the baseline regression. However, models in the GAMLSS setup may differ in the underlying distribution of the error terms. In the OLS setup the only distribution modelled is the Normal. In the GAMLSS there are families of distributions from which to choose. Once a distribution is chosen, models still differ because they may have different covariates.

The approach taken in this paper considers first choosing the distributions given a benchmark set of covariates. Table \ref{tab:selection} shows the Akaike information criteria (AIC) statistics for the estimated models and specifications for the underlying distributions of the errors. The AIC supports the truncated Weibull model with mean and scale shown in Tables \ref{tab:lossreg}. Table \ref{tab:gamlss1} in the appendix shows a set of models that are estimated. The key findings about the effect of GDP growth rates and the CPI are robust across models. When the dummy for the financial crisis is included, the effect of GDP growth becomes insignificant. However the model with crisis dummy is inferior in terms of the AIC.
\begin{table}[!ht]
\centering
\caption{Severity Model Selection}
  \label{tab:selection}
  \scalebox{0.9}{
  \begin{tabular}{ll}
  \toprule
    Distributions	                                                              &AIC\\
   \midrule
    Truncated Weibull (mean and scale with smoothing in regression)	            &-6,581\\
    Truncated generalized Pareto (mean and scale with smoothing in regression)	&-6,503\\
    Truncated Weibull (mean and scale regression)	                            &-6,362\\
    Truncated generalized Gamma (mean and scale regression)	                    &-6,361\\
    Truncated Weibull (only mean regression)	                                &-6,299\\
    Truncated generalized Gamma (only mean regression)	                        &-6,297\\
    \bottomrule
  \end{tabular}}
\end{table}

\noindent \textbf{Frequency regressions}

Table \ref{tab:freqreg} shows the result of the benchmark regression. GDP growth affects the number of annual loss events positively; more controls and more retail assets per employee reduce the number of losses. In the case of frequency, the CPI is not a significant regressor, neither a group of regressors such as government effectiveness, regulatory quality, rule of Law, control of corruption; all taken from the World Bank Worldwide Governance Indicators.

In addition, the regression in the scale or the variance of the number of losses shown in Table \ref{tab:freqreg} confirms that more employees per branch increase the variance in the number of annual losses and more operational risk controls reduce it.
\begin{table}[!ht]
\centering
\begin{threeparttable}
\caption{Regression Results in the Frequency Model}
  \label{tab:freqreg}
  \begin{tabular}{lllllr}
  \toprule
    Regressors	             &Estimates	 &Std. error  &t-value  &p-value & \\
    \midrule
    Regression in the mean   &           &            &         &        & \\
    \midrule
    Intercept	             &  4.33	&0.298	      &14.528	&0.000	&***\\
    GDP growth	             &  0.13	&0.019	      & 7.020	&0.000	&***\\
    Controls	             & -3.50	&0.575	      &-6.096	&0.000	&***\\
    Assets per employee	     & -0.03	&0.016	      &-1.969	&0.050	&.\\
    \midrule
    Regression in the scale   &           &            &         &        & \\
    \midrule
    Intercept	             & -0.14	 &0.507	       &-0.272	&0.786	& \\
    Employees per branch	 &  0.05	 &0.015	       & 3.019	&0.003	&**\\
    Controls (a)	         & -2.18	 &1.057	       & -2.064	&0.040	&*\\
    \bottomrule
  \end{tabular}
  \begin{tablenotes}
      \footnotesize{
      \item Note: Smoothing is performed with p splines and significance codes are $0 = ***$, $0.001=**$, $0.01=*$, $0.05=.$.}
    \end{tablenotes}
    \end{threeparttable}
\end{table}

Variants of the Poisson, Negative Binomial Type I, and Negative Binomial Type II \citep[see][]{rigby2017distributions} were estimated with GAMLSS. The chosen model, according to the AIC (see Table \ref{tab:selection2}) was the Negative Binomial with regressions in the mean and scale.

Table \ref{tab:gamlss2} in the appendix shows the details of the alternative regressions. Across all regressions, GDP growth stands robustly significant with a parameter value of 0.13. This means that there is a strong evidence that the opportunistic behavior described in Cressey's triangle also affects the number of fraud events. Hence, GDP growth affects both the number of events and the severity of those events. This is in line with the theory of opportunistic behavior of fraudsters when good aggregate economic times arrive.
\begin{table}[!ht]
\centering
\caption{Frequency Model Selection}
  \label{tab:selection2}
  \scalebox{0.9}{
  \begin{tabular}{ll}
  \toprule
     Distributions	                                                    & AIC\\
   \midrule
    Negative binomial (mean and scale with smoothing in regression)	    & 1,612\\
    Negative binomial II (mean and scale with smoothing in regression)	& 1,624\\
    Negative binomial (mean and scale regression)	                    & 1,632\\
    Negative binomial II (mean and scale regression)	                & 1,635\\
    Negative binomial (only mean regression)	                        & 1,637\\
    Negative binomial II (only mean regression)	                        & 1,659\\
 \bottomrule
  \end{tabular}}
\end{table}

\section{Conclusions} \label{sec:conclu}
This paper lays out a stochastic dynamic internal fraud model for operational risk to simulate operational losses due to internal fraud in retail banking. The model aims to capture the nature of internal fraud and the operational controls to mitigate or avoid the monetary losses caused by internal fraudsters. The model incorporates human factors such as the level of employees per branch as well as the ethical quality of workers. It also includes the endogenous risk controls.

The huge number of model parameters are calibrated by means of a shrinkage procedure that depends on available information about the heterogenous nature of banks in the ORX dataset. The simulated losses implied by the model mimic the aggregate ORX data in terms of the number of loss events and their severity in the retail banking, internal fraud cell as published in \citet{orx2012}.

Losses generated by the model are associated with GDP growth and the corruption perception of the country where banks are located. The generalized linear regression results are new because they uncover one important determinant of fraud losses not previously documented in the literature: The corruption perception index of a country. The results imply that higher corruption perception indices at the national level have a direct effect on the size of losses due to internal fraud events. The findings show that it is the yearly average size of losses that are affected by national corruption perceptions, not the frequency of losses within a year.

This paper can be extended in a number of ways. With real ORX data, the model parameters can be estimated with standard statistical procedures. Also, the causal effect from national corruption indicators to internal fraud in banks can be further explored. Currently, the paper does not make any distinction of losses originating in many countries within the international banks that have offices worldwide. The paper only considers the countries that have banks' headquarters.

The model described in the paper, with its benchmark calibration, can be also used to statistically compare data aggregation techniques dealing with consortium data. This can be done by means of monte-carlo simulations.

\pagebreak

\newpage
\bibliographystyle{chicago}
\bibliography{references}

\newpage
  \begin{center}
    {\Large{\textbf{Appendix}}}
  \end{center}
\appendix
\renewcommand{\thetable}{A-\arabic{table}}
\setcounter{table}{0}  

\section{Tables}\label{app:1}
\begin{table}[!ht]
\centering
 \begin{centering}
  \caption{Member Banks of the ORX Data Exchange by Country and Selected Dates}
  \label{tab:orxbanks}
\scalebox{.72}{
\begin{threeparttable}
\begin{tabular}{rlccccc}
\hline\hline
	&	\multirow{2}{*}{Name of bank}&	\multirow{2}{*}{country}&country &	bank 	&	membership	&	note	\\
    &                                &                          & code   &  code    &   date        & \\
\hline
1	&	Bank of Nova Scotia 	&	Canada   	                &	CAN	&	BNS	&	Apr-02	&		\\
2	&	Commerzbank AG 	        &	Germany	                    &	DEU	&	CBA	&	Apr-02	&		\\
3	&	Deutsche Bank AG 	    &	    Germany	                    &	DEU	&	DBA	&	Apr-02	&		\\
4	&	BNP Paribas 	        &	France	                    &	FRA	&	BNP	&	Apr-02	&		\\
5	&	ABN AMRO 	            &	Netherlands	                &	NLD	&	ABN	&	Apr-02	&		\\
6	&	Fortis NL	            &	Netherlands	                &	NLD	&	FTL	&	Apr-02	&	*	\\
7	&	ING Group 	            &	Netherlands	                &	NLD	&	ING	&	Apr-02	&		\\
8	&	JPMorgan Chase \& Co.	&	United States	            &	USA	&	JPM	&	Apr-02	&		\\
9	&	Bank of America	        &	United States	            &	USA	&	BOA	&	Mar-04	&		\\
10	&	West LB	                &	Germany	                    &	DEU	&	WLB	&	Jun-04	&	*	\\
11	&	Banesto	&	Spain	                                    &	ESP &	BNS	&	Jun-05	&	*	\\
12	&	SEB (Skandinaviska Enskilda Banken)	         &	Sweden	&	SWE	&   SEB &	Jun-05	&		\\
13	&	Credit Agricole SA 	   &	France	                    &	FRA	&	CAS	&	Dec-05	&		\\
14	&	Banc Sabadell	       &	Spain	                    &	ESP	&	BSB	&	Apr-06	&		\\
15	&	Cajamar	               &	Spain	                    &	ESP	&	CMR	&	Apr-06	&		\\
16	&	Barclays Bank	       &	United Kingdom	            &	GBR	&	BLB	&	Apr-06	&		\\
17	&	Bank Austria - Creditanstalt	    &	Austria	        &	AUT	&	BAC	&	Jun-06	&		\\
18	&	Fortis	               &	Belgium	                    &	BEL	&	FTS	&	Jun-06	&	*	\\
19	&	Caixa Catalunya	       &	Spain	                    &	ESP	&	CCT	&	Jun-06	&	*	\\
20	&	Banco Portugues de Negocios	        &	Portugal	    &	PRT	&	BPN	&	Jun-06	&	*	\\
21	&	National City	       &	United States	            &	USA	&	NAT	&	Jun-06	&	*	\\
22	&	Erste Group Bank AG    &	Austria	                    &	AUT	&	EGB	&	Ago-06	&		\\
23	&	BMO Financial Group    &	Canada	                    &	CAN	&	BMO	&	Oct-06	&		\\
24	&	Royal Bank of Canada (RBC) Financial Group 	&	Canada	&	CAN	&	RBC	&	Oct-06	&		\\
25	&	Banco Popular	       &	Spain	                    &	ESP	&	BPO	&	Dec-06	&	*	\\
26	&	Lloyds Banking Group   &	United Kingdom	            &	GBR	&	LBG	&	Dec-06	&		\\
27	&	US Bancorp	           &	United States	            &	USA	&	USB	&	Dec-06	&		\\
28	&	Grupo Santander	       &	Spain	                    &	ESP	&	BST	&	Jan-07	&		\\
29  &   Toronto Dominion Bank Group (TD BG)            & Canada &   CAN &   TDB &   Mar-07  &       \\
30	&	Banco Pastor	       &	Spain	                    &	ESP	&	BPS	&	Jun-07	&	*	\\
31	&	Caja Laboral	       &	Spain	                    &	ESP	&	CLB	&	Jun-07	&	*	\\
32	&	HBOS PLC	           &	United Kingdom	            &	GBR	&	HBO	&	Jun-07	&	*	\\
33	&	Wachovia Corporation   &	United States	            &	USA	&	WCR	&	Jun-07	&		\\
34	&	Washington Mutual	   &	United States	            &	USA	&	WAM	&	Jun-07	&		\\
35	&	PNC Bank	           &	United States	            &	USA	&	PNC	&	Nov-07	&		\\
36	&	Royal Bank of Scotland Group	    &	United Kingdom	&	GBR	&	RBS	&	Dec-07	&		\\
37	&	Bank of Ireland Group  &	Ireland	                    &	IRL	&	BIG	&	Dec-07	&		\\
38	&	HSBC Holdings plc	   &	United Kingdom	            &	GBR	&	HSB	&	Jan-08	&		\\
39	&	Hana Bank	           &	South Korea	                &	KOR	&	HBK	&	Jan-08	&	*	\\
40	&	Rabobank Nederland	   &	Netherlands	                &	NLD	&	RBN	&	Jan-08	&		\\
41	&	National Australia Bank	&	Australia	                &	AUS	&	NAB	&	Sep-08	&		\\
42	&	Banco Bradesco S/A 	   &	Brazil	                    &	BRA	&	BSC	&	Sep-08	&		\\
43	&	Caixanova	           &	Spain	                    &	ESP	&	CNV	&	Sep-08	&	*	\\
44	&	Wells Fargo \& Co	   &	United States	            &	USA	&	WFC	&	Sep-08	&		\\
45	&	First Rand 	           &	South Africa	            &	ZAF	&	FRD	&	Sep-08	&		\\
46	&	Deutsche Postbank AG   &	Germany	                    &	DEU	&	DPB	&	Nov-08	&		\\
47	&	Standard Chartered Bank	                 &	Singapore	&	SGP	&	STA	&	Dec-08	&		\\
48	&	Capital One	           &	United States	            &	USA	&	CON	&	Mar-09	&		\\
49	&	Bank of New York Mellon            &	United States	&	USA	&	BNY	&	Jul-09	&		\\
50	&	Westpac Banking Corporation 	   &	Australia	    &	AUS	&	WBC	&	Nov-09	&		\\
51	&	Commonwealth Bank of Australia	   &	Australia	    &	AUS	&	CBA	&	Dec-09	&		\\
52	&	Soci\'et\'{e} G\'{e}n\'{e}rale 	   &	France	        &	FRA	&	SGL	&	Dec-09	&		\\
\hline\hline
\end{tabular}
  \begin{tablenotes}
      \footnotesize{
      \item Note: * marks banks that are not members of the ORX association any more. }
    \end{tablenotes}
\end{threeparttable}}
\end{centering}
\end{table}
\noindent \textbf{Notes on Table \ref{tab:orxbanks}}:

The main source of the list of banks in Table \ref{tab:orxbanks} is \citet{orx2016}. The list only shows banks that were active members of the ORX association in the period 2006-2010. As shown, member banks are gathered all over the world but belong mainly to advanced economies. Some members banks do not belong to the association anymore because they filed into bankruptcy especially as a consequence of The Global Financial Crisis of 2007-2009 or because they were absorbed by other banks.

The main source is then complemented by partial listings of ORX membership appearing on a number of presentations \citep[see][]{Patel2009, Sabatini2008, Sabatini2009, Kennett2010}

Given the information in Table \ref{tab:orxbanks}, the model calibration implied a varying number of total banks in the sample. For example, up to end of 2008, there are N=39 banks. In year 2009, 10 banks entered the data exchange association, and four banks quit the association, making $N=45$ banks participating in the data exchange. By the end of 2010, one more banks entered the association, making $N=48$ active banks. We track 52 banks in total and 35 banks that belongs to ORX the entire five-year period. The key workable assumption is that arrivals and departures from the association are set at the beginning of each year. In addition, the database contained only banks that operated a retail-banking segment. Therefore, some banks that belong to the ORX association but only perform investment banking or other lines of business were omitted from the database.

\newpage

\begin{table}[!ht]
\scalebox{0.9}{
\begin{threeparttable}
\caption{Severity regressions with GAMLSS}
  \label{tab:gamlss1}
  \begin{tabular}{lcccccc}
  \toprule
    Mean regression   &Model 1	   &Model 2	   & Model 3	&Model 4	&Model 5	& Model 6\\
    Intercept	      &-2.80	   & -2.78	   &  -2.43	    & -2.65	    & -2.47	    & -2.59\\
	                  & (***)	   & (***)	   &  (***)	    &  (***)	&  (***)	& (***)\\
    GDP growth	      & 0.01	   &	       &   0.01	    &  0.01	    &  0.01	    &  0.01\\
	                  & (*)		   &           &    (*)	    &   (*)	    &   (*)	    &  (*)\\
    Crisis		      &            & -0.08	   &			&           &           &     \\
        		      &            &  (**)	   &		    &           &           &     \\
    CPI	              &-0.05	   & -0.05	   &  -0.17    	& -0.12     & -0.16	    & -0.10\\
	                  & (***)	   &  (***)	   &  (***)	    &  (***)	&  (***)	&  (.)\\
    Government Effectiveness&	   &		   &   0.33		&	        &           & \\
			          &            &           &  (***)		&	        &           &\\
    Regulatory Quality&			   &	       &            &  0.21		&           &\\
				      &            &           &            &   (*)		&           &\\
    Rule of Law		  &			   &           &            &           &  0.29	    &\\
					  &            &           &            &           &  (***)	&\\
    Control of Corruption&		   &		   &		    &           &           &   0.10\\
					  &	           &           &            &           &           &\\
    Control	          &-0.40	   &  -0.63	   &   -0.26	& -0.44	    & -0.42	    &  -0.42\\
	                  & (*)	       & (***)	   &	        &   (*)	    &  (*)	    &   (*)\\
    Employees per branch& 0.13	   &   0.14	   &    0.13	&  0.14	    &  0.14	    &   0.13\\
	                    & (***)	   &  (***)	   &    (***)	&  (***)	& (***)	    & (***)\\
    Assets per employee	&-0.01	   &  -0.01	   &    -0.01	& -0.00	    & -0.00	    &  -0.01\\
	                    &  (.)	   &   (*)	   &     (*)    & 	                    &  (***)\\
    \midrule
    Scale  regression &            &           &            &           &          & \\
    \midrule
    Intercept	      &   0.73	   &  0.56	   &   1.12	    &  0.83	    &  1.12	   &    0.92\\
	                  &   (***)	   &  (***)	   &   (***)	&  (***)	&  (***)   &   (***)\\
    GDP growth	      &   0.01	   &  -0.02	   &			&           &          & \\
	                  &    (*)	   &		   &		    &           &          &\\
    CPI	              &  -0.07	   &  -0.02	   &   -0.20	& -0.15	    & -0.21	   &  -0.11\\
	                  &   (***)	   &   (.)	   &   (***)	&  (***)	&  (***)   &  (***)\\
    Employees per branch&  -0.01   &  -0.01	   &   -0.02	& -0.02	    & -0.02	   &  -0.01\\
	                    & (***)	   &   (***)   &   (***)	&  (***)	&  (***)   &  (***)\\
    \midrule
    Diagnostics         &          &           &            &           &          & \\
    \midrule
    Global Deviance	   & -6657.35  &-6633.58   &-6677.04	&-6718.08	&-6732.36  &-6658.17\\
    AIC	               & -6581.01  &-6553.02   &-6598.11	&-6610.79	&-6624.83  &-6577.47\\
    SBC	               & -6344.62  &-6303.55   &-6353.68	&-6278.53	&-6291.84  &-6327.55\\
    \bottomrule
  \end{tabular}
  \begin{tablenotes}
      \footnotesize{
      \item All models include additional regressors that are defined in terms of smoothed terms. They are not reported here because they are used as additional controls. Smoothing is performed with p-splines.
      \item Significance codes are $0 = '***'$, $0.001='**'$, $0.01='*'$, $0.05='.'$, $0.1=''$.}
    \end{tablenotes}
    \end{threeparttable}}
\end{table}

\newpage

\begin{table}[!ht]
\scalebox{0.85}{
\begin{threeparttable}
\caption{Frequency regressions with GAMLSS}
  \label{tab:gamlss2}
  \begin{tabular}{llllllll}
  \toprule
    Mean regression      &Model 1	&Model 2	& Model 3  & Model 4	&Model 5	& Model 6  & Model 7\\
    Intercept	         &  4.33	&  4.32	    &  4.28	   &  4.23	    &  4.22	    &  4.24	   &  4.40\\
	                     & (***)	&  (***)	&  (***)   &  (***)	    & (***)	    &  (***)   &  (***)\\
    GDP growth	         &  0.13	&  0.13	    &  0.13	   &  0.13	    &  0.14	    &  0.14	   &  0.12\\
	                     & (***)	&  (***)	& (***)	   &  (***)	    & (***)	    &  (***)   &  (***)\\
    Dummy for crisis     &	        &  0.04		&		   &	        &           &          & \\
					     &		    &           &          &            &           &          &\\
    CPI			         &          &           &  0.02	   &			&           &          &\\
					     &		    &           &          &            &           &          &\\
    Gov. effectiveness   &			&	        &          &  0.12		&	        &          &\\
					     &		    &           &          &            &           &          &\\
    Regulatory Quality   &			&		    &          &            &  0.13		&          &\\
					     &		    &           &          &            &           &          &\\
    Rule of Law		     &			&	        &          &            &           &  0.17	   &\\
					     &		    &           &          &            &           &          &\\
    Control of Corruption&			&			&	       &            &           &          &  0.00\\
						 &          &	        &          &            &           &          &\\
    Control	             & -3.50	& -3.51	    & -3.68	   & -3.70	    & -3.69	    & -3.83	   & -3.69\\
	                     &(***)	    &  (***)	& (***)	   &  (***)	    &  (***)	& (***)	   &  (***)\\
    Assets per employee	 & -0.03	& -0.03	    & -0.03	   & -0.03	    & -0.03	    & -0.04	   & -0.04\\
	                     &  (.)	    &  (*)	    &  (*)	   &   (*)	    &   (*)	    &  (*)	   &    (*)\\
    \midrule
    Scale  regression    &          &           &          &            &           &          & \\
    \midrule
    Intercept	         & -0.14	& -0.14	    & -0.11	   & -0.10	    & -0.07	    & -0.07	   & -0.47\\
						 &	        &           &          &            &           &          & \\
    Employees per branch &	0.05	& 0.05	    &  0.05    &  0.05	   &  0.05	    &  0.05	   &  0.05\\
	                     &  (**)	& (**)	    &  (**)    &  (**)    &  (**)	    &  (**)	   &  (**)\\
    \midrule
    Diagnostics          &          &           &          &            &           &          &\\
    \midrule
    Global Deviance	     & 1581.63	& 1581.58	& 1570.44  & 1581.11	& 1570.75	& 1562.64	& 1535.25\\
    AIC	                 & 1611.76	& 1613.72	& 1606.82  & 1612.27	& 1609.40	& 1604.36	& 1584.57\\
    SBC	                 & 1663.01	& 1668.40	& 1668.74  & 1665.30	& 1675.15	& 1675.34	& 1668.48\\
    \bottomrule
  \end{tabular}
  \begin{tablenotes}
      \footnotesize{
      \item 1) All models include additional regressors that are defined in terms of smoothed terms. They are not reported here because they are used as additional controls. Smoothing is performed with p-splines.
      \item 2) Significance codes are $0 = '***'$, $0.001='**'$, $0.01='*'$, $0.05='.'$, $0.1=''$.}
    \end{tablenotes}
    \end{threeparttable}}
\end{table}

\renewcommand{\thefigure}{B-\arabic{figure}}
\setcounter{figure}{0}  
\pagebreak[4]
\section{Figures}\label{app:2}
\begin{figure}[!ht]
 \centering
 \caption{Distribution of parameters of the ramp function that defines the outbreak of operational losses in each bank.}
 \label{fig:app1}
 \begin{tabular}{cc}
   \includegraphics[scale=.3]{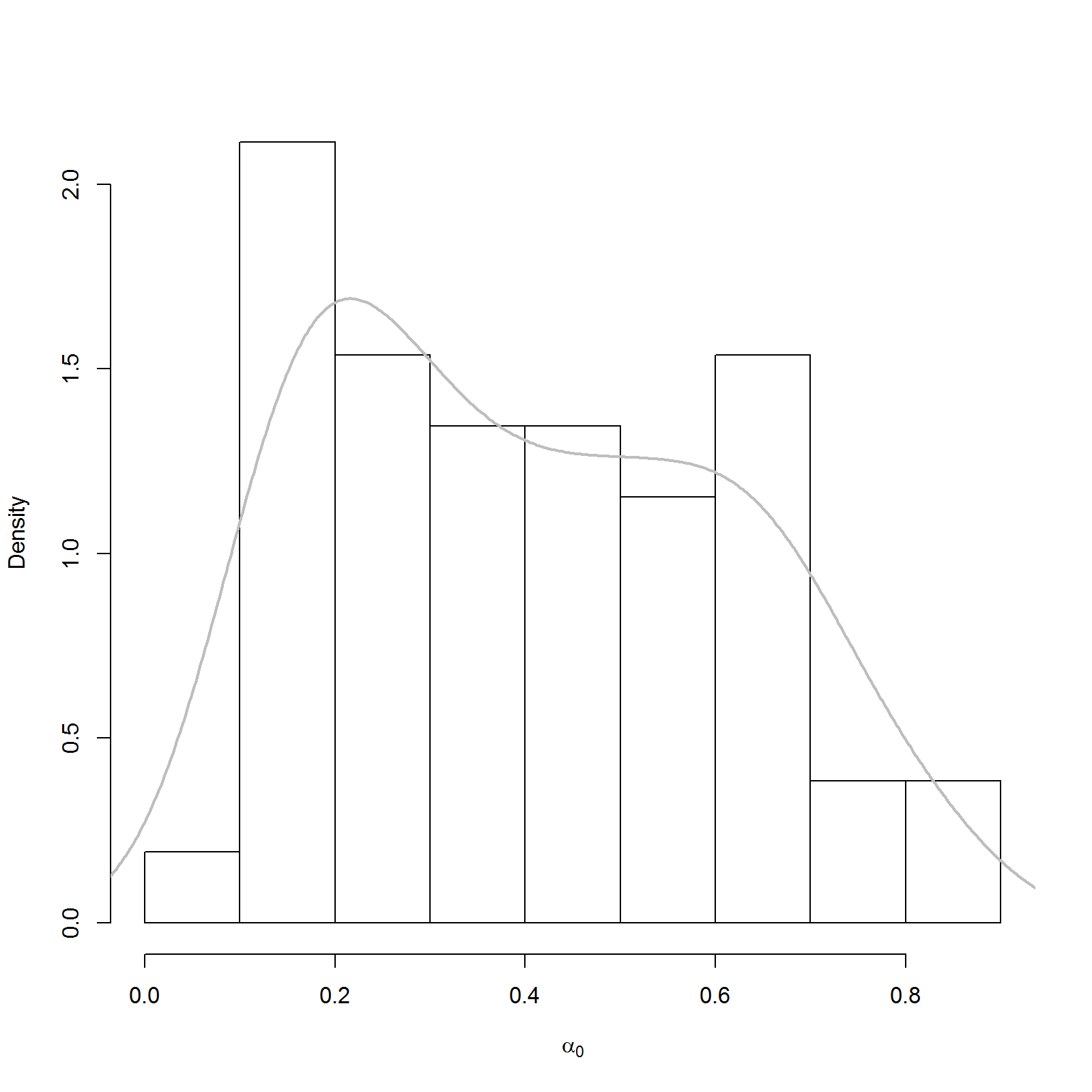} & \includegraphics[scale=.3]{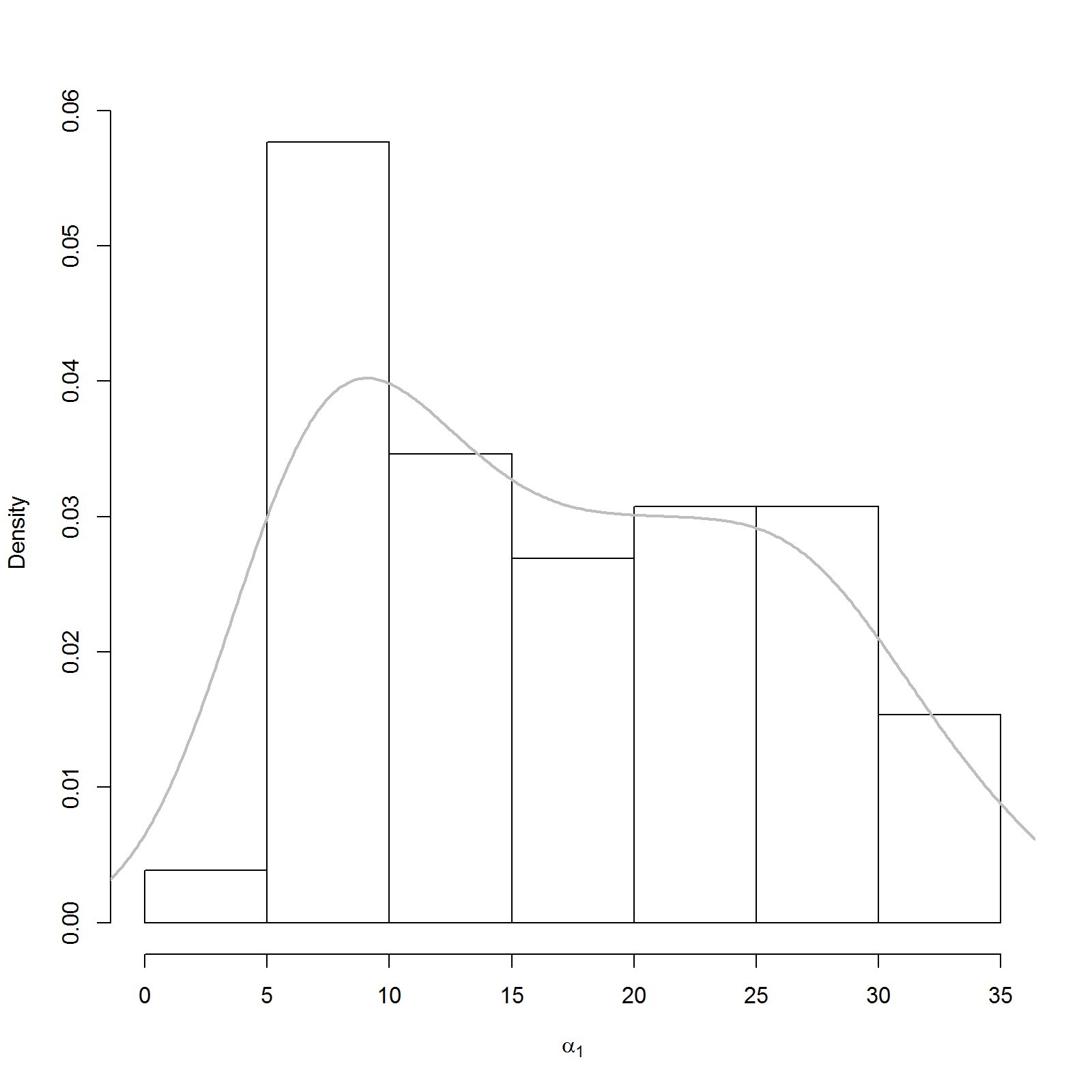} \\
   \includegraphics[scale=.3]{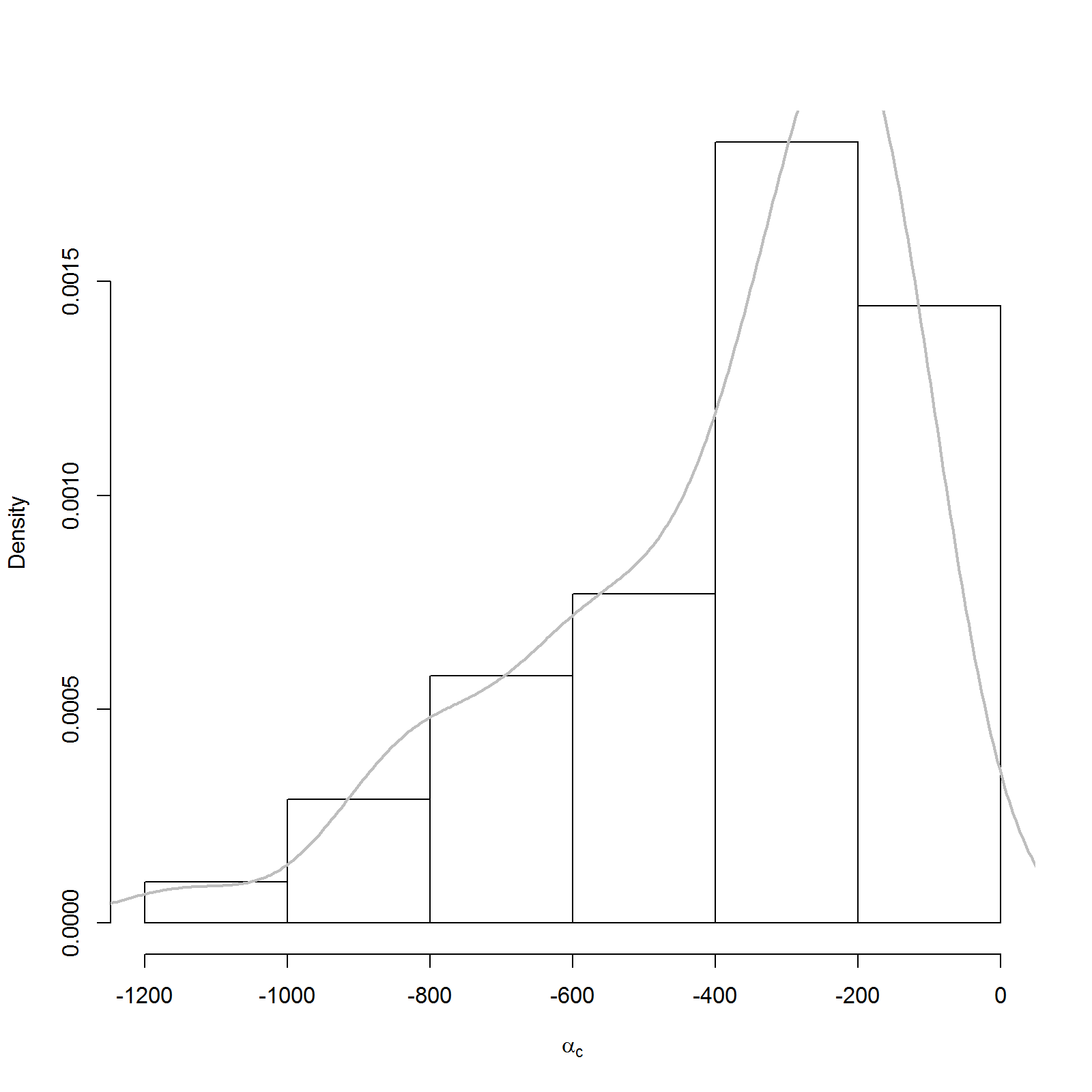} & \includegraphics[scale=.3]{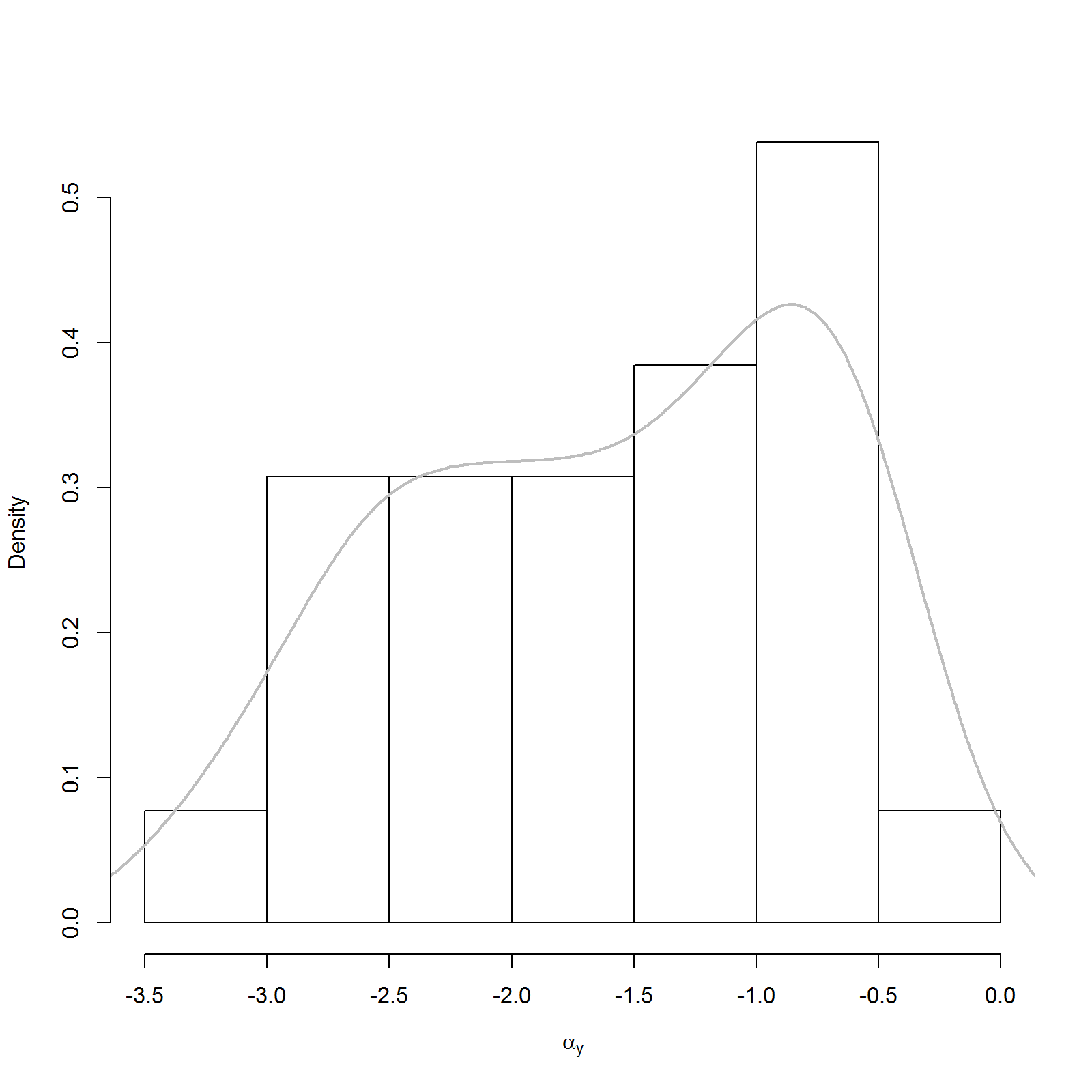} \\
   \includegraphics[scale=.3]{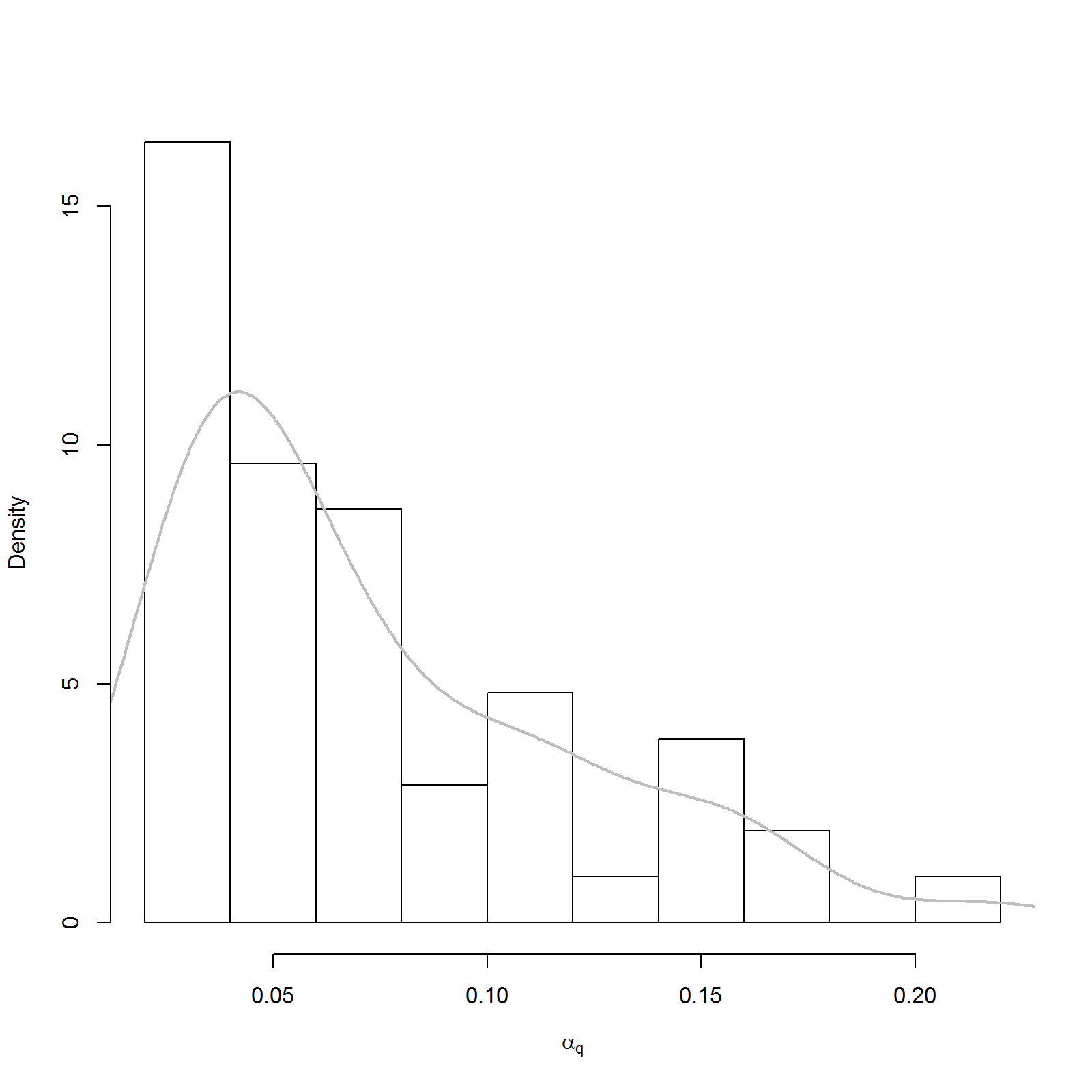} &  \\
 \end{tabular}
\end{figure}

\begin{figure}[!ht]
 \centering
 \caption{Distribution of other idiosyncratic parameters in the operational loss model.}
 \label{fig:app2}
 \begin{tabular}{cc}
   \includegraphics[scale=.3]{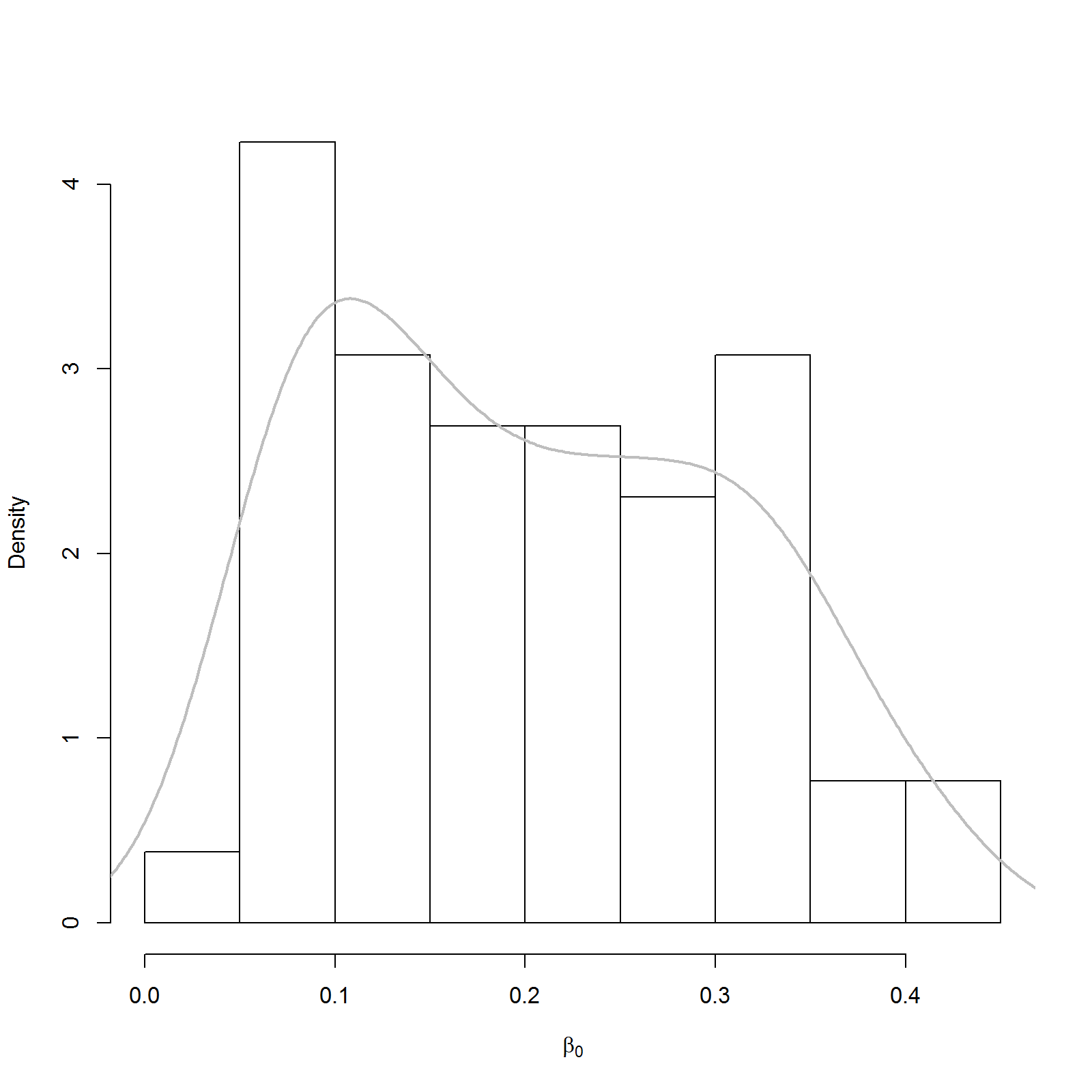} & \includegraphics[scale=.3]{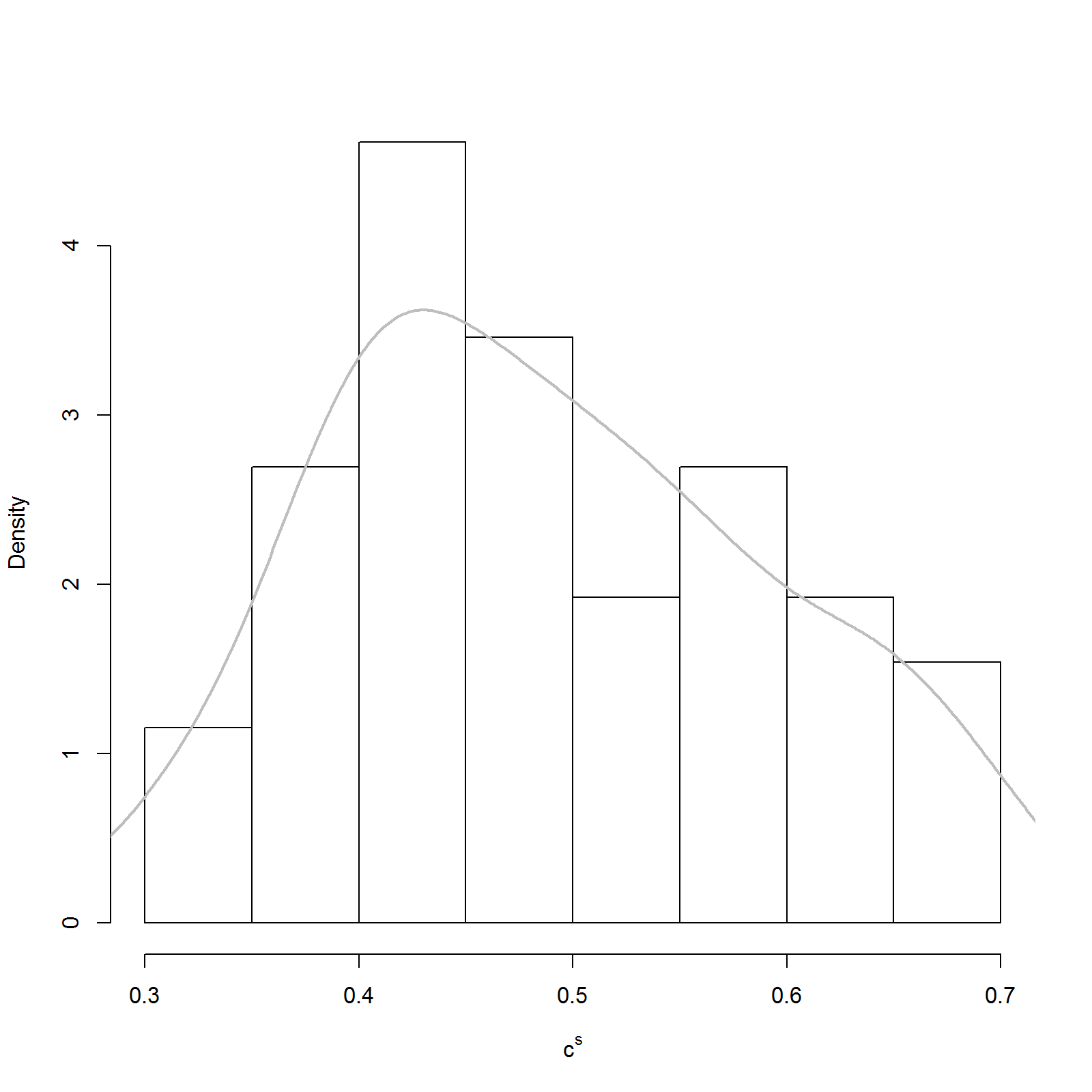} \\
   \includegraphics[scale=.3]{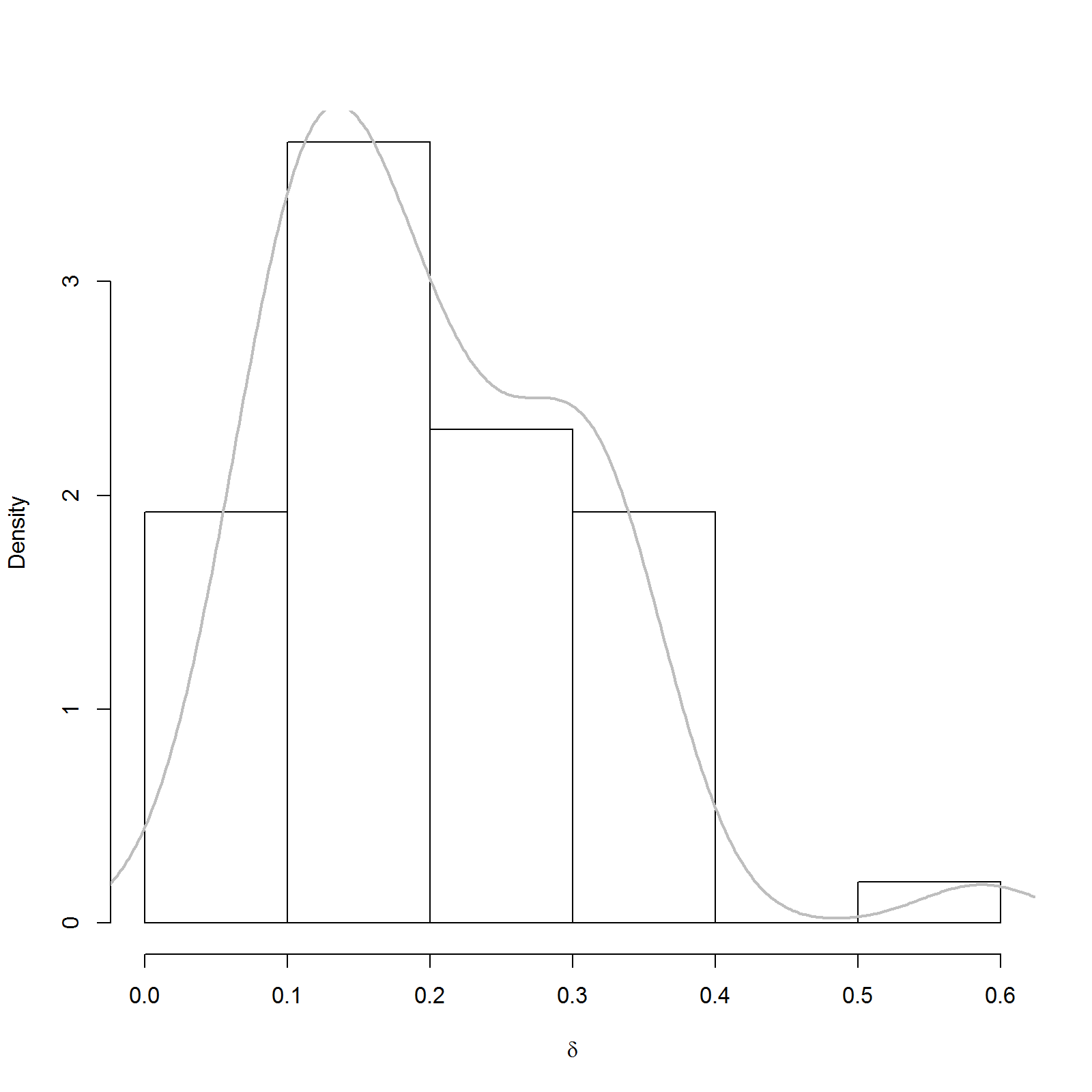} & \includegraphics[scale=.3]{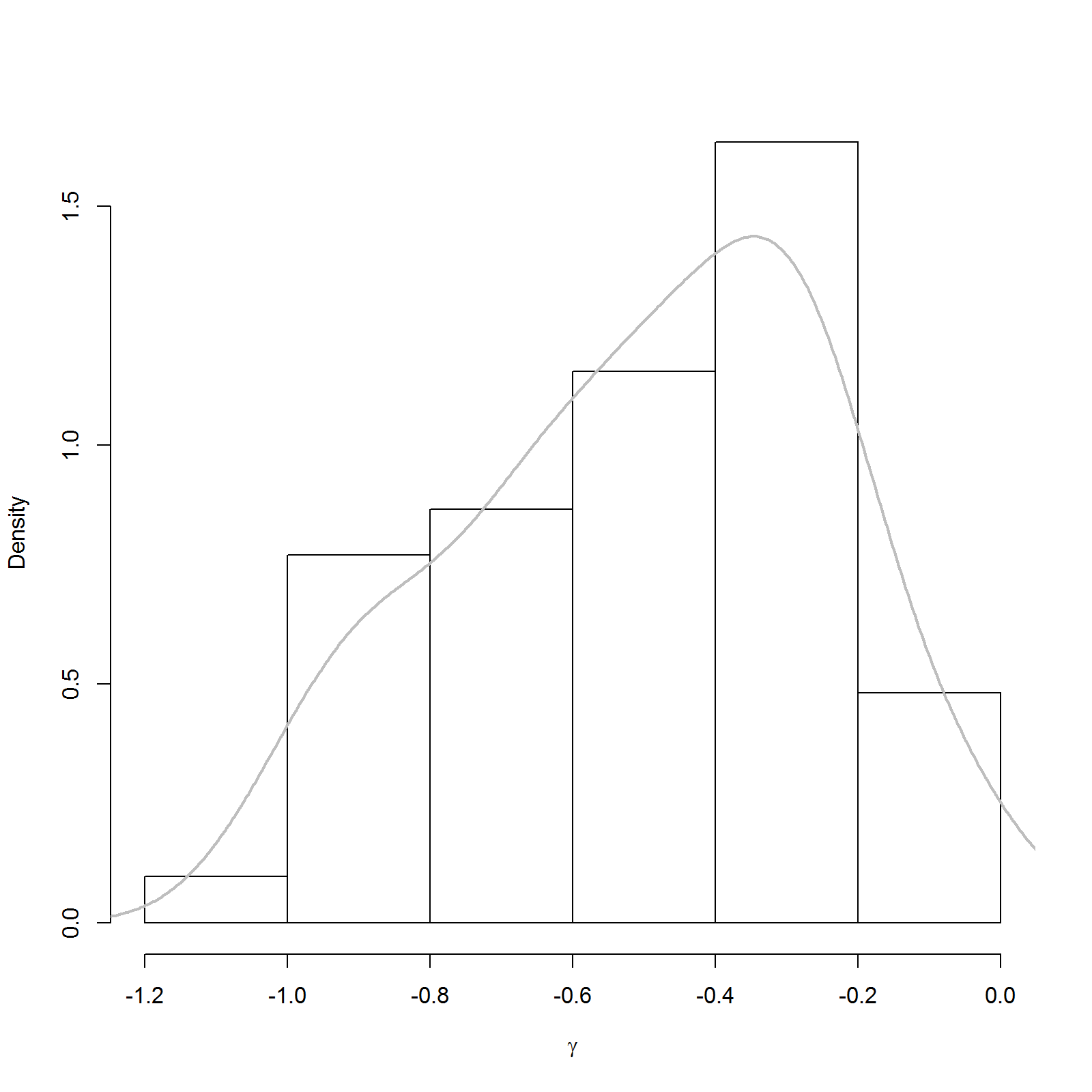} \\
   \includegraphics[scale=.3]{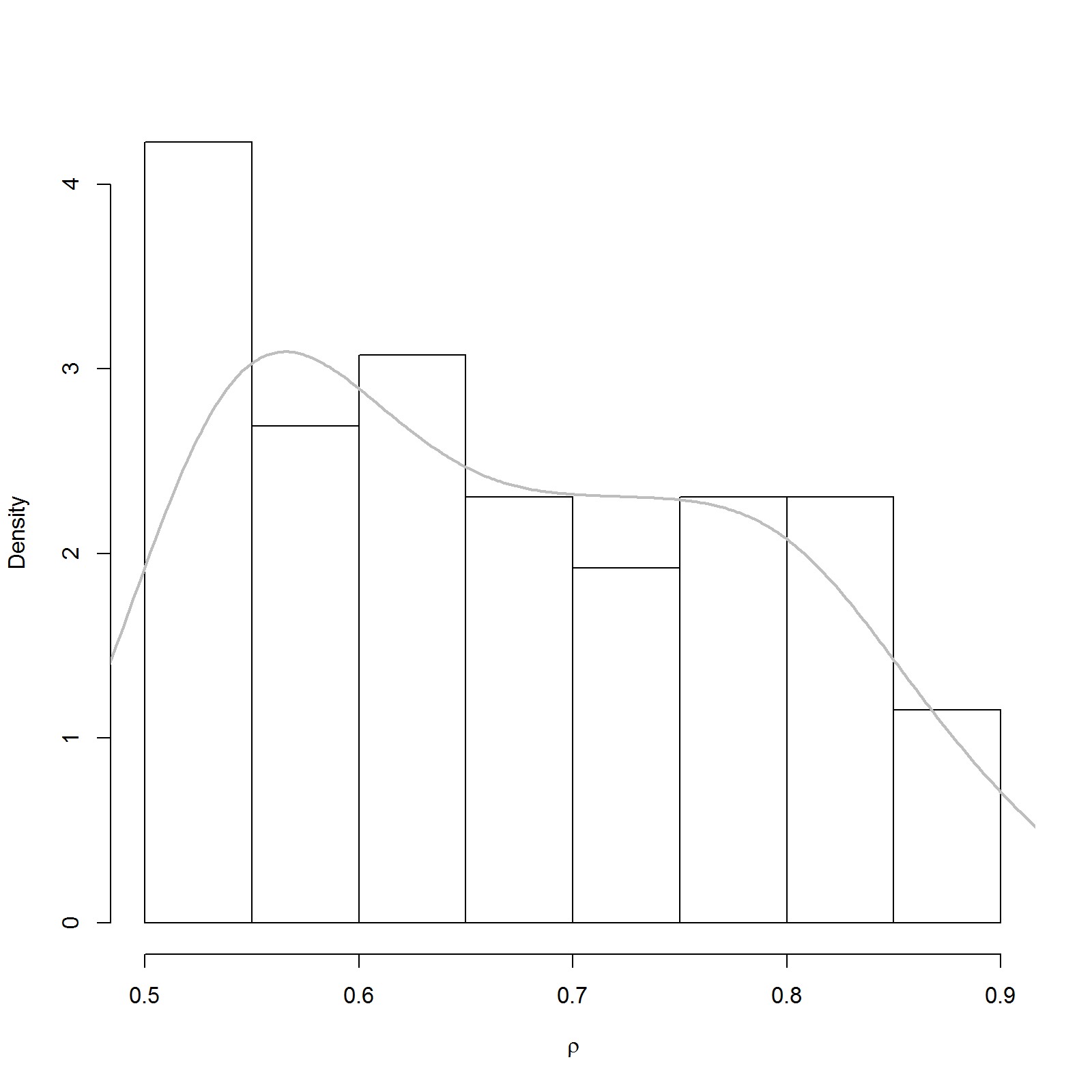}   & \includegraphics[scale=.3]{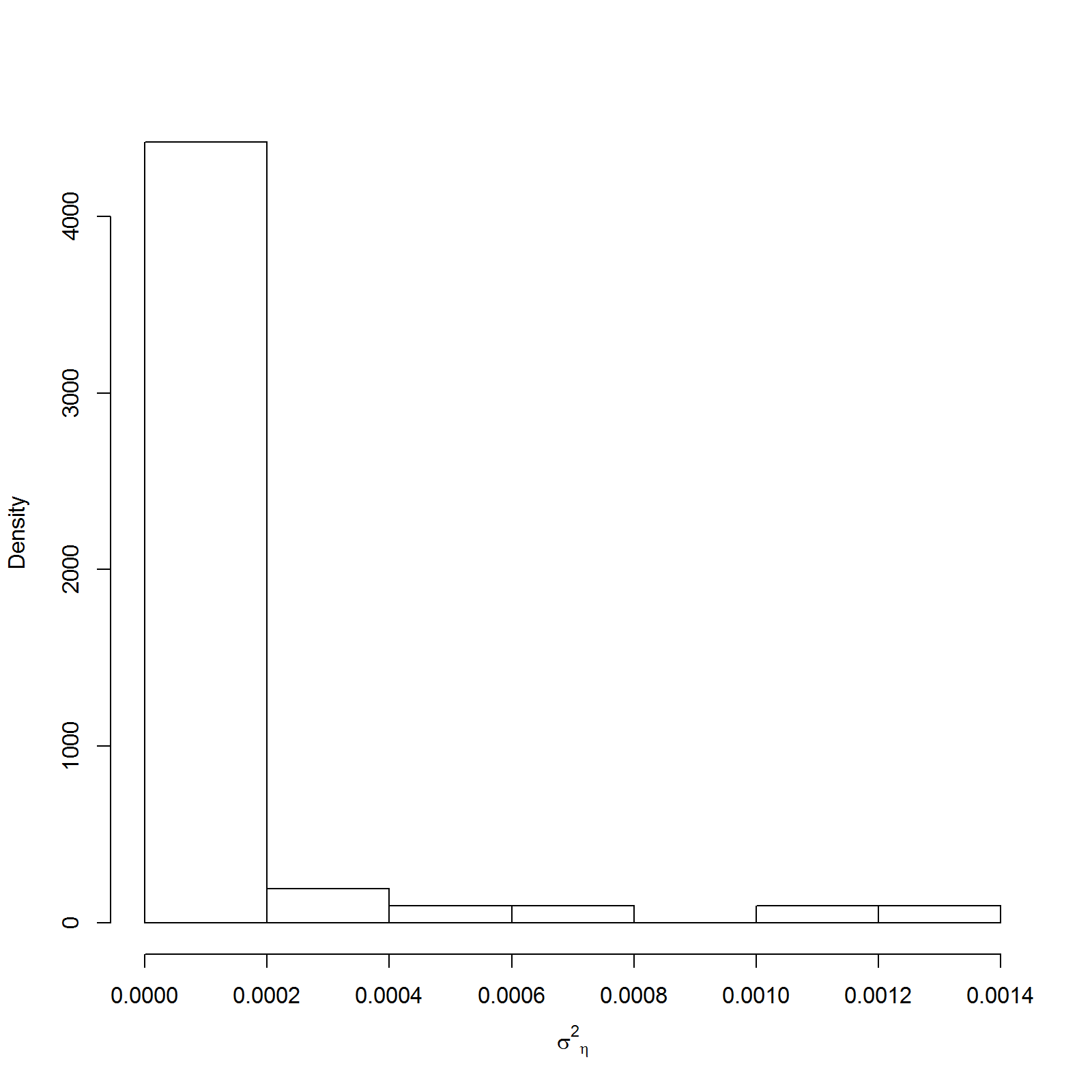} \\
 \end{tabular}
\end{figure}

\end{document}